\documentclass[twocolumn,tighten]{aastex63}

\usepackage{amsmath}
\usepackage{xspace}

\usepackage[encapsulated]{CJK}
\usepackage{ucs}
\usepackage[utf8x]{inputenc}
\newcommand{\cntext}[1]{\begin{CJK}{UTF8}{gbsn}#1\ignorespacesafterend\end{CJK}}  

\newcommand{\keyw}[1]{\textcolor{gray}{#1}}
\newcommand{\change}[1]{\textcolor{black}{#1}}

\newcommand{\kabs}{\ensuremath{\kappa_\nu^{\mathrm{abs}}}\xspace}

\newcommand{\kabst}{\ensuremath{\kappa_\nu^{\mathrm{abs,tot}}}\xspace}
\newcommand{\kscat}{\ensuremath{\kappa_\nu^{\mathrm{sca,tot}}}\xspace}

\newcommand{\amin}{\ensuremath{a_\mathrm{min}}\xspace}
\newcommand{\amax}{\ensuremath{a_\mathrm{max}}\xspace}

\usepackage{longtable}
 
\begin{document}
\title{Self-consistent ring model in protoplanetary disks: temperature dips and substructure formation} 

\correspondingauthor{Shangjia Zhang}
\email{shangjia.zhang@unlv.edu}

\author[0000-0002-8537-9114]{Shangjia Zhang\cntext{(张尚嘉)}}
\affiliation{Department of Physics and Astronomy, University of Nevada, Las Vegas, 
                4505 S. Maryland Pkwy, Las Vegas, NV, 89154, USA}
                
\author[0000-0003-3201-4549]{Xiao Hu\cntext{(胡晓)}}
\affiliation{Department of Physics and Astronomy, University of Nevada, Las Vegas, 
                4505 S. Maryland Pkwy, Las Vegas, NV, 89154, USA}
\affiliation{Department of Astronomy, University of Virginia, 530 McCormick Road, Charlottesville, VA 22904, USA}

\author[0000-0003-3616-6822]{Zhaohuan Zhu\cntext{(朱照寰)}}
\affiliation{Department of Physics and Astronomy, University of Nevada, Las Vegas, 4505 S. Maryland Pkwy, Las Vegas, NV, 89154, USA}

\author[0000-0001-7258-770X]{Jaehan Bae}
\altaffiliation{NASA Hubble Fellowship Program Sagan Fellow}
\affil{Earth and Planets Laboratory, Carnegie Institution for Science, 5241 Broad Branch Road NW, Washington, DC 20015, USA}

\begin{abstract}

Rings and gaps are ubiquitous in protoplanetary disks.
Larger dust grains will concentrate in gaseous rings more compactly due to stronger aerodynamic drag.
However, the effects of dust concentration on the ring's thermal structure have not been explored. Using MCRT simulations, we self-consistently construct ring models by iterating the ring's thermal structure, hydrostatic equilibrium, and dust concentration. We set up rings with two dust populations having different settling and radial concentration due to their different sizes. We find two mechanisms that can lead to temperature dips around the ring. When the disk is optically thick, the temperature drops outside the ring, which is the shadowing effect found in previous works adopting a single-dust population in the disk. When the disk is optically thin, a second mechanism due to excess cooling of big grains is found. Big grains cool more efficiently, which leads to a moderate temperature dip within the ring where big dust resides. This dip is close to the center of the ring.
Such temperature dip within the ring can lead to particle pile-up outside the ring and feedback to the dust distribution and thermal structure. We couple the MCRT calculations with a 1D dust evolution model and show that the ring evolves to a different shape and may even separate to several rings.  Overall, dust concentration within rings has moderate effects on the disk's thermal structure, and self-consistent model is crucial not only for protoplanetary disk observations but also for planetesimal and planet formation studies. 
\end{abstract}


\section{Introduction \label{sec:intro}}

Recent high angular resolution observations of protoplanetary disks have revealed many substructures. 
\citet{ andrews20} summarizes them into four categories: \textit{rings/cavity} (which are ``transition'' disks with bright rings and depleted cavities, e.g., LkCa 15, J1610 \citealt{facchini20}, GM Aur \citealt{huang20}, PDS 70 \citealt{keppler19}), \textit{rings/gaps} (which are concentric, axisymmetric patterns of enhancing and depleting intensity, e.g.,  HL Tau \citealt{ALMA2015}, TW Hya \citealt{andrews12}, RU Lup \citealt{andrews18b}, AS 209 \citealt{guzman18}), \textit{arcs} (which are non-axisymmetric substructures with a partial ring extending only a certain azimuthal angle, e.g., MWC 758 \citealt{dong18a}, HD 163296 \citealt{isella18b}) and \textit{spirals} (ranging from m=2 to asymmetrical spirals, e.g., IM Lup, Elias 27 \citealt{huang18c}). Among them, the first two kinds are observed most frequently. We call them \textit{rings} hereafter.

Dozens of ring forming mechanisms have been proposed, including tidal interaction between the planet and the disk \citep{lin79, goldreich80}, disk dispersal with MHD-driven winds \citep{takahashi18} or photoevaprative flows \citep{ercolano17},  zonal flows \citep{johansen09} in MHD disks, mass pile up at the boundary between magnetically  active and dead zones \citep{flock15}, spontaneous ring formation due to reducing accretion by concentrated dust \citep{dullemond18a, hu2019}, and condensation fronts at icelines \citep{zhang2015}.

All these mechanisms except icelines generate gaseous pressure bumps which trap  dust grains. Intermediate-sized particles with Stokes number of about unity drift fastest responding to gaseous bumps. Small grains (St $\ll$ 1) are well-coupled to the gas, whereas very big grains (St $\gg$ 1) are fully decoupled. Under the protoplanetary disk condition at 100 au, cm particles have the most significant concentration within the gaseous bumps. There are indeed some tentative results from multi-wavelengths observations which indicate that grains are larger (towards cm size) at the ring and smaller (towards mm size) in the gap \citep{macias19, carrasco2019, huang20, long20}. 

Different sized grains have different concentration on the vertical direction too. 
Very small grains ($<$1$\mu$m) have similar scale height as that of the gas, and are best probed by near-infrared scattered light observations. Bigger grains are settled to the midplane, and probed by (sub)mm/cm radio observations. The vertical extent of the settled grains depends on the strength of turbulence in the disk. The vertical settling of the grains is balanced by the turbulent diffusion that stirs up these grains \citep{youdin2007}. Stronger turbulence leads to thicker dusty layers. This vertical dust diffusion applies to the whole disk and also to rings.

These effects of dust concentration should also affect ring's thermal structure, since dust opacity is the primary source of opacity in the disk. 
The disk is heated by stellar radiation and viscous heating, and cooled predominantly by dust thermal emission. The disk temperature is set when heating balances cooling. Studying disk temperature self-consistently is important for interpreting observations (e.g., the decrease of temperature in shadow can be misinterpreted as density depletion, \citealt{isella2018}), and understanding disk dynamics including the Vertical Shear Instability \citep{Nelson2013} and baroclinic instability \citep{klahr03}. Furthermore, the temperature structure determines the radial pressure gradient (dP/dr) which directly affects dust trapping itself. This means that the thermal structure and dust concentration could have a feedback loop on each other: dust concentration changing the temperature structure while the temperature structure affecting the dust consternation. It is unclear if such feedback loop can result in a stable or unstable disk configuration.  Thus, it is essential to consider dust distribution and disk thermal structure self-consistently.

In this work, we construct self-consistent models of rings by using two dust species with different density distributions. We iterate the ring's thermal structure, hydrostatic equilibrium and dust concentration with MCRT calculations. In Section \ref{sec:toymodel}, we use a toy model to demonstrate that the varying opacity in the ring can lead to moderate temperature variation, in addition to the previously studied shadowing effect. In Section \ref{sec:ringmodel}, we present our setup for the systematic study of the shadowing and the opacity effects, using one and two populations of dust grains. In Section \ref{sec:dustevomodel}, we  couple MCRT with a 1D dust evolution model, and demonstrate that the feedback loop can change the shape of the initial ring. 
In Section \ref{sec:discussion}, we propose that the temperature gap in a recent observation can be more naturally explained by the opacity effect. We also discuss some other observational perspectives there. We conclude our work in Section \ref{sec:conclusion}.

\section{A toy model for excess cooling \label{sec:toymodel}}
Due to dust trapping by a pressure bump, the dust inside and outside a gaseous ring have different size distributions, which leads to different opacities inside and outside the ring.  
Since the opacity affects the disk temperature (e.g., \citealt{calvet91}), it is very likely that the temperatures inside and outside the ring are different.

For an optically thin disk region which  has only absorption opacity ($\kappa_{\nu}^{abs}$), the heating $Q_+$ comes from the stellar irradiation at all wavelengths,
\begin{equation}
    Q_+ = \int_{0}^{\infty} \kabs F_\nu^* d\nu.
\end{equation}
If we assume that the star radiates as a black body, we have
\begin{equation}
    F_{\nu}^{*} = \frac{4\pi R_*^2 \pi B_\nu(T_*)}{4\pi r^2},
\end{equation}
where $R_*$ is the stellar radius and $T_*$ is the stellar effective temperature.

The disk cools by its own radiation at the disk temperature ($T_d$) which is much lower than the stellar temperature. The cooling term is
\begin{equation}
    Q_- = 4\pi \int_{0}^{\infty} \kabs B_\nu(T_d) d\nu.
\end{equation}
When the disk is in thermal equilibrium, 
$Q_+ = Q_-$.

If we use the Planck mean opacity
\begin{equation}
    \kappa_P^{abs}(T) \equiv \frac{\int_{0}^{\infty} \kabs B_\nu(T)d\nu}{\int_0^{\infty} B_\nu(T)d\nu},
\end{equation}
we can derive the disk temperature using $Q_+ = Q_-$ and express the disk temperature as
\begin{equation}
    T_d = \bigg(\frac{R_*}{2r}\bigg)^{\frac{1}{2}} \frac{1}{\epsilon^{1/4}} T_*,
\end{equation}
where $r$ is the distance to the star and
\begin{equation}
    \epsilon \equiv \frac{\kappa_P(T_d)}{\kappa_P(T_*)},
    \label{eq:coolcoeff}
\end{equation}
is the thermal cooling coefficient. For the gray opacity whose $\kappa_{\nu}^{abs}$ is a constant with $\nu$, $\epsilon$ is unity. This $\epsilon$ parameter is essential for determining the disk's vertical structure \citep{calvet91,chiang97}.
Considering that $B_{\nu}(T_d)$ peaks at mm wavelengths, we can approximate $\kappa_{P}(T_d)$ using the monochromatic opacity at 1 mm, $\kappa(\lambda = 1 mm)$. For the same reason, we have $\kappa_{P}(T_*)\approx\kappa(\lambda = 1 \mu m)$. Thus, we have
$\mathrm{\epsilon \approx \kappa(\lambda = 1 mm)/\kappa(\lambda = 1 \mu m)}$. 

For very big grains, the opacity is mostly gray, so  $\epsilon \approx 1$. For small grains, the opacity at $\mu$m is larger than the opacity at mm wavelengths, so $\epsilon < 1$. Hence, the equilibrium temperature for the disk with mainly small grains is higher than the temperature for the disk mainly with big grains. Compared with small grains, big grains cool more efficiently with respect to the amount of energy they absorbed. 
Since there are much more big grains inside the ring than outside the ring, the temperature in the ring is expected to be lower.

To demonstrate this effect, we start with a power-law opacity ($\kabs \propto \nu^{\beta}$) in a disk. The disk temperature can be solved as,
\begin{equation}
    T_d = \bigg(\frac{R_*}{2r}\bigg)^{\frac{1}{2}\frac{1}{1+\beta/4}} T_*.
\end{equation}
Thus, a disk with smaller grains (having a larger $\beta$) is hotter than a disk with bigger grains. When both small and big grains are present in one disk, the disk temperature at one radius will be determined by the dominant dust species at that radius.  

We set up a ring and use the MCRT code RADMC-3D \footnote{\url{https://www.ita.uni-heidelberg.de/~dullemond/software/radmc-3d/}} to calculate the temperature. For simplicity, we use a low density disk and only consider the absorption opacity. These simplifications highlight the opacity effect. (The more realistic setup will be given in the next section.)

\begin{figure*}[t!]
\includegraphics[width=\linewidth]{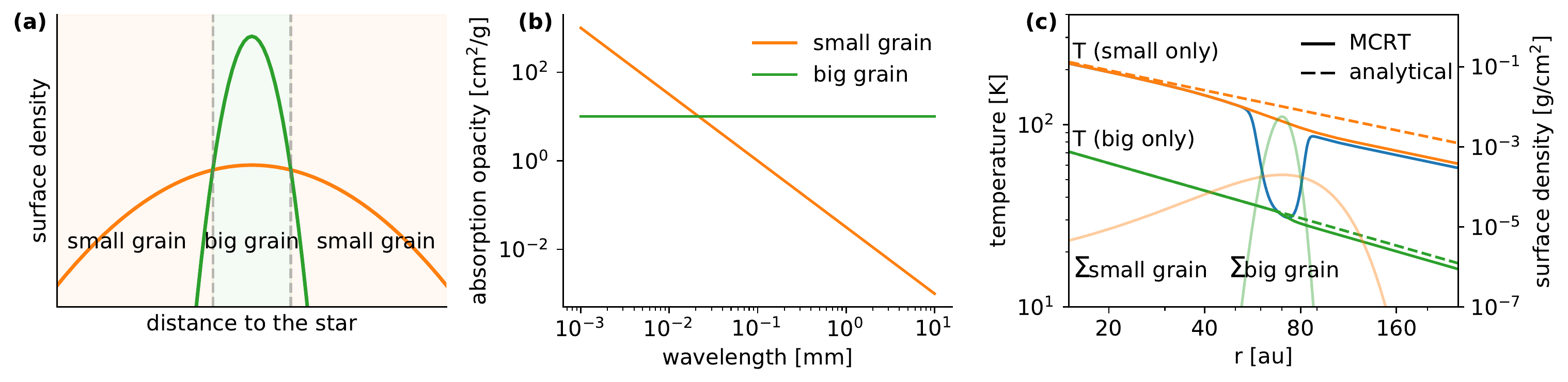}  
\figcaption{(a) The surface density of the toy model. The small (big) grain is represented by the orange (green) curve. The big grain has a narrower width and dominates inside the ring (green), whereas the small grain dominates outside the ring. (b) The absorption-only opacity used for the toy model. The big grain is represented by a constant opacity, whereas the small grain is represented by a opacity $\propto \nu^{1.5}$. (c) The temperature calculations. The orange (green) curves show temperature if the disk only has small (big) grains. The solid lines are calculations from RADMC-3D MCRT, and the dashed lines are analytical solution assuming optical thin. The  small grain has a higher equilibrium temperature at the outer disk. The MCRT result when including both species is represented by the blue curve. The temperature approaches the small grain's outside the ring and approaches the big grain's inside the ring. Their respective surface density profiles are marked in transparent colors.
\label{fig:demo}}
\end{figure*}

The ring has two different dust species and their radial distributions are shown in Figure \ref{fig:demo} (a). Both species have Gaussian surface density, and the big grains have larger peak surface density, but less radial and vertical extension due to radial trapping and vertical settling \citep{dullemond18b}. The opacity for the small and big grains have slopes $\beta$ = 0 and 1.5, respectively (Figure \ref{fig:demo} b). The Orange curves in the rightmost panel are the temperatures from a disk which only has small grains, and the green curves are the temperatures from a disk which only has big grains. The temperatures from the MCRT calculations (solid orange and green curves) follow their analytical solutions (dashed lines) closely. The slight deviation of the temperature starting from the ring location indicates that the optically thin assumption is no longer valid. The blue curve is the temperature from a disk where both dust species are considered in the MCRT calculation. It approaches the small-grain-only temperature out of the ring and big-grain-only temperature within the ring (Figure \ref{fig:demo} c). This is in a good agreement with the expectation that small grains dominate the opacity far away from the ring center while big grains dominate the opacity close to the ring center. 
 
\begin{figure*}[t!]
\includegraphics[width=\linewidth]{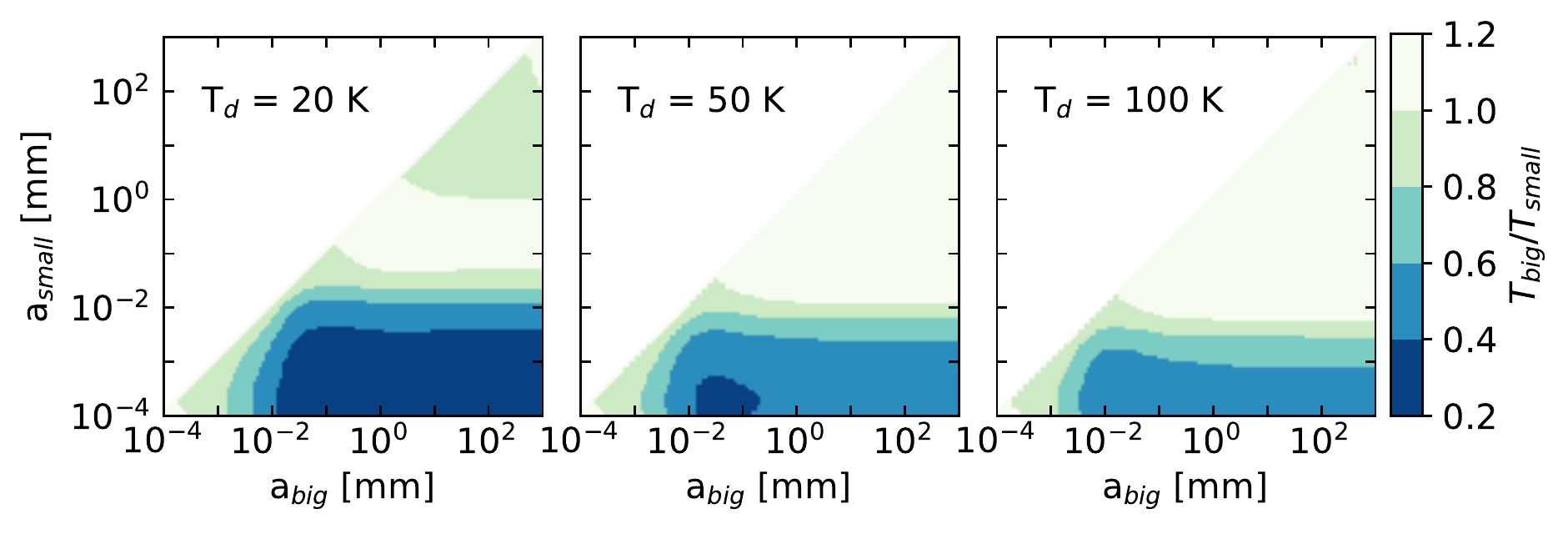}  
\figcaption{The equilibrium temperature ratio between the single-sized big species and the small species calculated using DSHARP opacity. The stellar temperature is 6000 K. From left to right, the disk temperatures T$_d$ are 20, 50 and 100 K. 
\label{fig:singlesize2d}}
\end{figure*}

\begin{figure}[t!]
\includegraphics[width=\linewidth]{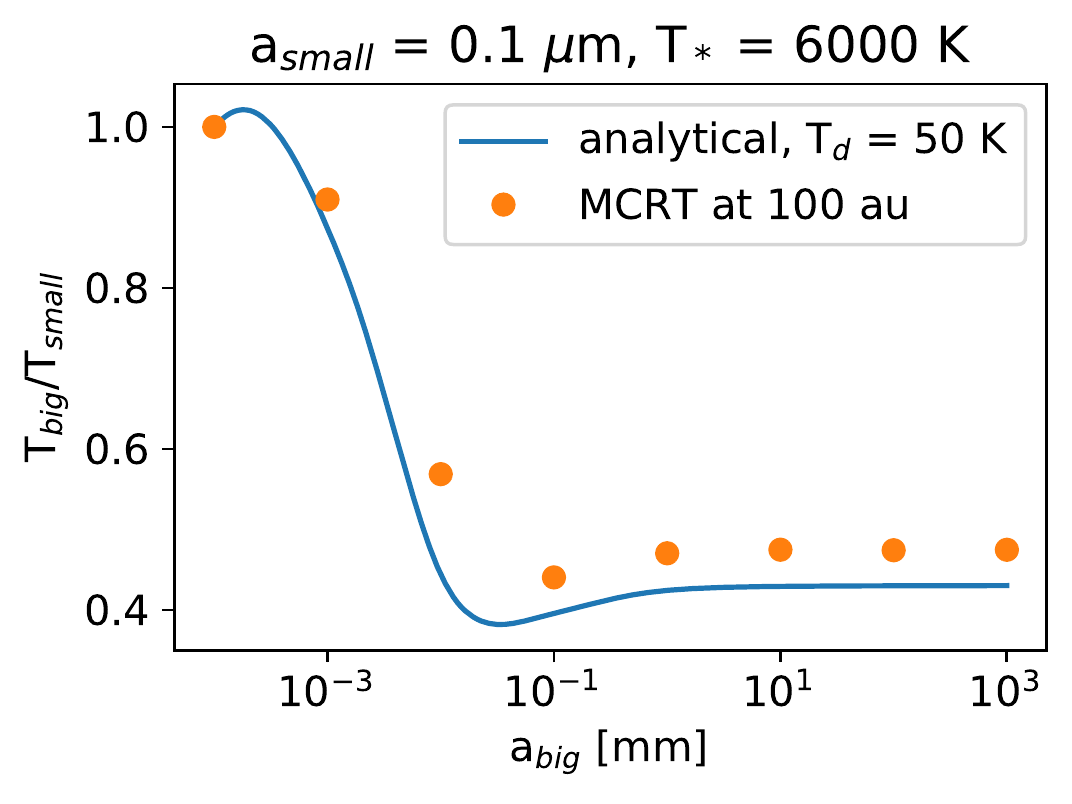}  
\figcaption{Comparison of analytical result in Figure \ref{fig:singlesize2d} with MCRT at 100 au, in the optically thin limit. 
\label{fig:singlesize2dcomp}}
\end{figure}

In the optically thin limit, the ratio between the temperature inside and outside the ring is the ratio between big-grain-only and small-grain-only equilibrium temperature, T$_{big}$/T$_{small}$. This ratio is ($\epsilon_{small}/\epsilon_{big}$)$^{1/4}$ and can be calculated using Equation \ref{eq:coolcoeff}. We use single-sized DSHARP opacity \citep{birnstiel18} and plot the temperature ratio for various pairs of small and big grains in Figure \ref{fig:singlesize2d}. Different disk temperatures have been explored. The ratio is smaller for a lower disk temperature (i.e., outer disk), since these parts of the disk emit at longer wavelengths leading to smaller $\epsilon_{small}$. For disk temperature at 20 K, the ratio can be as low as 25\% if the big grain is around hundreds of $\mu$m and the small grain is around 0.1 $\mu$m, which can be a realistic situation in protoplanetary disks. In a hotter disk (e.g., at the inner disk), the ratio can be as low as 50\%. On the anti-diagonal lines, the ratio is unity when the big and small grains have the same size. To validate these analytical estimates, we calculate a smooth disk temperature using RADMC-3D. We adopt single sized opacities, with a$_{small}$ = 0.1$\mu$m and various a$_{big}$. The temperature ratios measured in MCRT at 100 au are dotted in Figure \ref{fig:singlesize2dcomp} along with an analytical curve picked in Figure \ref{fig:singlesize2d} with the T$_d$ = 50 K, where $T_{big}$ and $T_{small}$ are measured from two different single-population simulations at 100 au. The analytical result predicts the Monte Carlo calculation closely.

Even though we only have two single-sized dust populations in Figure \ref{fig:singlesize2d}, these contours should still roughly apply to more realistic dust size distributions where the opacity is normally dominated by the biggest dust particles. 

\section{Ring models \label{sec:ringmodel}}
In this section, we explore the temperature structure with ring configurations in more realistic disk configurations. We adopt a more realistic opacity (Section \ref{sec:opacity}), and iterate the dust and thermal structure of the disk (Section \ref{sec:thermal}). We consider models with one-population (Section \ref{sec:onepop}) or two-population (Section \ref{sec:twopop}) dust species in the disks.

\subsection{Opacity \label{sec:opacity}}

\begin{figure}[t!]
\includegraphics[width=\linewidth]{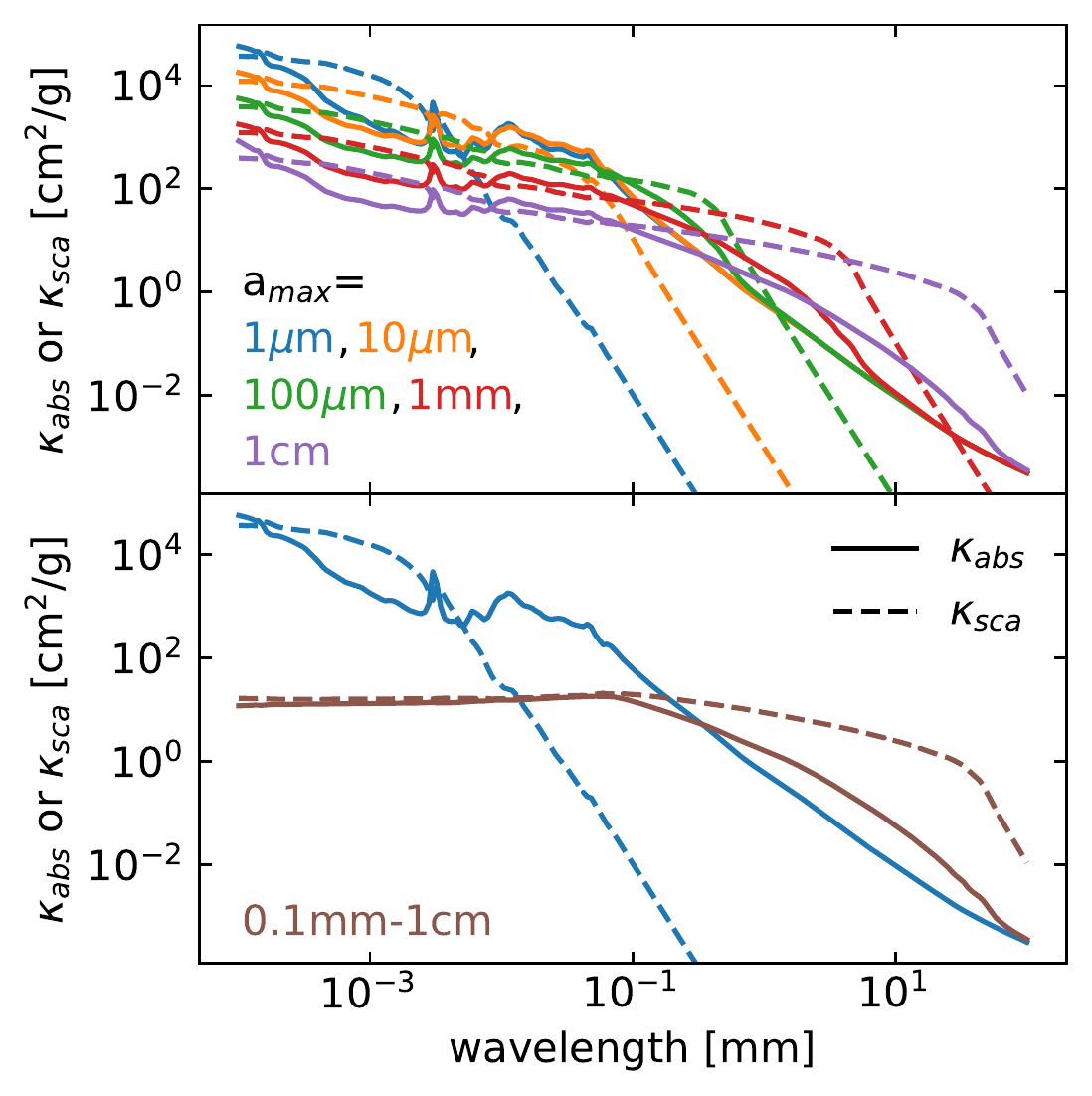}  
\figcaption{Top: The dust opacities used for the disk with a single dust population. Optical constants are from DSHARP \citep{birnstiel18}. The minimum grain size is 0.1 $\mu$m.  At short wavelengths, the opacity decreases as the \amax increases. Blue, orange, green, red and purple denote \amax = 1 $\mathrm{\mu}$m, 10 $\mathrm{\mu}$ 100 $\mathrm{\mu}$m, 1 mm and 1 cm cases, respectively. Solid curves represent absorption opacities (\kabst) and dashed curves represent scattering opacities (\kscat). Bottom: The opacities used for disks with two dust populations. The opacities between 0.1mm-1cm are marked in red-brown.
\label{fig:opacity_all}}
\end{figure}

We assume a power-law dust size distribution with maximum particle size \amax, minimum particle size \amin, and power index $q$. The number density of particles follows,
\begin{align}
 n(a) \propto \, \begin{cases}
 a^{-q} & \text{for} \, a_\mathrm{min} \leq a \leq a_\mathrm{max} \\
 0      & \text{else}.                                            
 \end{cases}
 \label{eq:powerlaw}
\end{align}
We fix $q=3.5$ in this work, so the dust mass is top-heavy and proportional to $a^{0.5}$. The optical constants are taken from the DSHARP opacity in \citet{birnstiel18} (see also references therein). The opacity is calculated using the package \texttt{dsharp\_opac} \citep{github_dsharp_opac}\footnote{\url{https://github.com/birnstiel/dsharp_opac}}. For the one-population models and the small grain population in two-population models, we adopt \amin = 0.1 $\mu$m. For one-population setups, we use opacities with \amax = 1, 10, 100$\mu$m, 1mm and 1cm (see the top panel of Figure \ref{fig:opacity_all}). \change{For two-population setups, we use opacities from \{$\ensuremath{a_\mathrm{min, small}}$, $\ensuremath{a_\mathrm{max, small}}$\} = \{0.1 $\mu$m, 1 $\mu$m\} to represent small grains and \{$\ensuremath{a_\mathrm{min, big}}$, $\ensuremath{a_\mathrm{max, big}}$\} = \{0.1 mm, 10 mm\} to represent big grains (see the bottom panel of Figure \ref{fig:opacity_all}). }

Note that in two-population models, the opacities used in small grain population and big grain population follow the power-law size distribution with $q=3.5$, respectively. However, the combination of these two populations has a dust size distribution varying spatially, since the small and big grains have different scale heights and radial widths (see also in Section \ref{sec:twopop}). In our configuration, we calibrate the surface density ratio between dust (including both populations) and gas as 1:100 only at the Gaussian ring's peak. Likewise, the surface density ratio between two populations only follows a power-law with $q=3.5$ at the Gaussian ring's peak. This ratio is ($\ensuremath{a_\mathrm{max, big}}^{0.5} - \ensuremath{a_\mathrm{min, big}}^{0.5}$)/($\ensuremath{a_\mathrm{max, small}}^{0.5} - \ensuremath{a_\mathrm{min, small}}^{0.5}$)=31.6\footnote{\change{$\ensuremath{a_\mathrm{max, small}}$ used here is 10 $\mu$m instead of 1 $\mu$m. This gives the small-grain population a larger mass fraction. Otherwise, the ratio is around 132. We adopt the former ratio, but this does not affect results in this paper.}}. Since the small grain population has a larger vertical scale height, the volume density does not follow this ratio even at the ring's peak. The surface density ratio between two populations is smaller than 31.6 away from ring's peak, since the small grain population has larger ring width. The dust to gas mass ratio is less than 1:100 away from the ring's peak.

\subsection{Self-consistent thermal structure \label{sec:thermal}}
We setup gas densities assuming vertical hydrostatic equilibrium. We distribute the dust based on the dust surface density and vertical scale height (see Section \ref{sec:onepop} and \ref{sec:twopop}). Then,
we calculate the temperature using MCRT code RADMC-3D. Initially, the gas, small and big grains are assumed to be at the same temperature. This is the case if small grains are coupled with the gas, and big and small grains reach thermal equilibrium through collisions. The new derived temperature from MCRT will be used to calculate a new vertical hydrostatic equilibrium structure. These processes are iterated until a converged solution has been reached.

In detail, to initialize the MCRT simulations, we assume that the disk is vertically isothermal and assign a Gaussian density profile in the vertical direction. The midplane temperature is set as
\begin{equation}
    T_{irr} (r) = \Big(\frac{fL_*}{4\pi R^2\sigma_{SB}} \Big)^{1/4},
\end{equation}
where $f$ accounts for the flaring of the disk, and $f= 0.05$ in our initialization. $L_*$ is the stellar luminosity. We adopt $L_*$ = 1.05 $L_\odot$ here. $\sigma_{SB}$ is the Stefan-Boltzmann constant. This temperature is used to calculate the scale height $h$ = $c_s/\Omega_K$ (we adpot $M_*$ = 1.25 $M_\odot$ for $\Omega_K$). Then, we run RADMC-3D to get the $R-\theta$ temperature distribution for the first iteration. At this point, the new temperature is higher at the disk surface and is not vertically isothermal anymore. Thus, the vertical hydrostatic equilibrium solution needs to be adjusted. We use the hydrostatic equation in the vertical direction to calculate a new density profile,
\begin{equation}
    -\frac{GM_*z}{(r^2 + z^2)^{3/2}} - \frac{1}{\rho}\frac{\partial P}{\partial z} = 0,
\end{equation}
where $G$ is the gravitational constant, z is the height, $\rho$ is the gas volume density. $P = \rho RT/\mu$ is the gas pressure. $R$ and $\mu$ are the gas constant and mean molecular weight (we adopt $\mu$=2.4). The vertically integrated surface density is constrained to be the profile we prescribe. Then we put this new density into RADMC-3D and iterate these steps for several times until the temperature is converged. We use five iterations in all of our models. The detail of temperature iteration can also be found in \citet{bae19} Appendix A.

For all runs, we use $10^8$ photons and assume isotropic scattering. The calculation is in $R-\theta$ plane with mirroring boundary condition (in one quadrant of the meridian plane).  The radial grid is 512 in logarithmic scale and vertical $\theta$ grid is 128 from 0 to 24$^\circ$ above the midplane. The inner boundary is 1 au and the outer boundary is 300 au. We assume the star as a point source, since our interests are at the outer disk midplane ($R_* \ll r$)  where the size of the star is unimportant. The temperature of the star is 4500 K.

\subsection{One-population models \label{sec:onepop}}
As a first step and a baseline model for our two-population models, we run ring models with a single dust population. In these cases, the dust size distribution does not vary spatially.  Nevertheless, shadowing effect can still affect the midplane temperature distribution. The mechanism can be understood as the following. The stellar radiation is intercepted by a puffed-up region in the disk. This region casts a shadow to the outer disk, so the temperature behind the puffed-up region drops. Since the disk flares, it comes out of the shadow at farther distance, and becomes brighter again. Thus, in the radial direction, the temperature first decreases and then increases. The idea of self-shadowing by a disk inner rim was first introduced to explain the observations of Herbig Ae/Be stars \citep{dullemond04a}.  Later studies focus on the planet gap carved by planets \citep{jang-condell12, jang-condell13, isella2018}. This effect can lead to a temperature contrast as high as $\sim$ 20\%, depending on the planet mass (or the gap's shape). The contrast here refers to the deviation from a smooth temperature profile with a smooth surface density profile.

\begin{figure*}[t!]
\includegraphics[width=\linewidth]{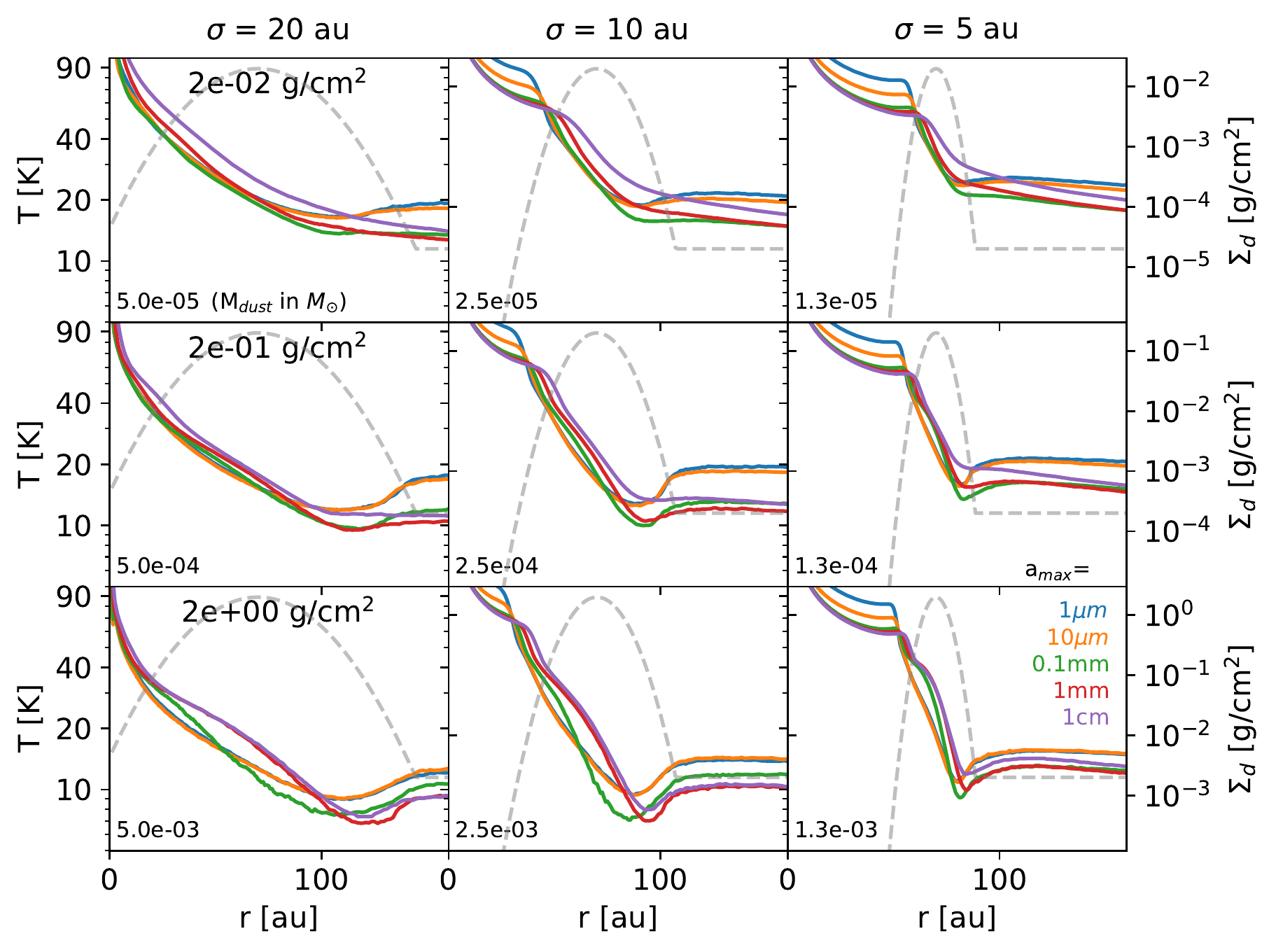}  
\figcaption{The midplane temperature of one population models with ring configurations. From top to bottom, the dust surface density at the Gaussian peak are 0.02, 0.2 and 2 $\mathrm{g\, cm^{-2}}$. From left to right, the Gaussian widths are 20, 10 and 5 au. $\amax$ with increasing sizes are marked with blue, orange, green, red and magenta lines. The dashed lines are the dust surface density in logarithmic scale. The floor density over the peak density is $10^{-3}$. The total dust mass is shown in the bottom left corner.
\label{fig:1pop}}
\end{figure*}

We carry out a systematic parameter study for the rings. We put a Gaussian ring at 70 au. The ring width $\sigma$ = 5, 10 and 20 au. The dust surface density at the Gaussian peak $\Sigma_{peak}$ = 0.02, 0.2 and 2 $\mathrm{g\, cm^{-2}}$. We put a density floor at the outer disk, so that the dust is not completely depleted.  The floor density over the peak density $\Sigma_{floor}/\Sigma_{peak}$ = 0.1, 0.01, $10^{-3}$ and $10^{-4}$. The radial profile can be described as the following,
\begin{align}
\Sigma_{d}(r) = \, \begin{cases}
 \Sigma_{peak}\text{exp} \Big( -\frac{(r - 70\ \textrm{au})^2}{2\mathrm{\sigma^2} }  \Big)
  \ \text{if} \ \Sigma_{d} > \Sigma_{floor},  \\
 \Sigma_{floor} \ \text{if} \ \Sigma_{d} \leq \Sigma_{floor}\ \text{and}\ r>r_{peak}.                                            
 \end{cases}
 \label{eq:gaussianbump}
\end{align}
These models are summarized in Table \ref{table}. We find that for this shadowing effect, the level of density floor matters, since it defines the sharpness of a ring. We will discuss this issue in Section \ref{sec:discussion}. Here, we first study the effects from other parameters, using $\Sigma_{floor}/\Sigma_{peak}$ = 0.001. The midplane temperatures of models with different surface density profiles are shown in Figure \ref{fig:1pop}. From top to bottom the density increases. The width of the ring decreases from left to right. In all models, the resulting temperature is not a power-law across the ring region. Instead, the temperature dip occurs at the outer edge of the ring. With the same density setup, the temperature profile depends on the opacity. In most of the cases, the temperature dip is stronger  when $\amax$ = 1 or 10 $\mu$m. The temperature variation respective to a smooth temperature profile can reaches $\sim$25\%. These values are roughly consistent with previous works focusing on the temperature profiles around the planetary gaps \citep{jang-condell12, jang-condell13, isella2018}. 
The temperature profile is smoother for $\amax$ = 1 cm cases, since grains with larger $\amax$ have lower opacity at short wavelengths and higher opacity at long wavelengths. They emit more efficiently thus have lower equilibrium temperatures at the outer disk. These disks are less puffed-up. As the density increases or the ring becomes wider, the temperature becomes lower. This is also because the cooling is more efficient with more dust. A narrow ring also leads to a larger temperature gradient dT/dr. 


The shadowing effect happens when the disk is optically thick. As the disk becomes more optically thick, the temperature dip becomes deeper. To quantify this, we calculate the Rosseland mean opacities of different dust size distributions. If the disk temperature T = 20K, the Rosseland mean absorption opacity $\kappa_{R, abs}$ = 9.27, 9.44, 16.8, 20.8, 8.40 $\mathrm{cm^{2}\, g^{-1}}$ for \amax = 1$\mu$m, 10$\mu$m, 100$\mu$m, 1mm and 1cm. $\tau$=1 happens when the dust surface density at the peak $\Sigma_{peak}$ $\approx$ 0.1 $\mathrm{g\, cm^{-2}}$ for \amax = 1$\mu$m, 10$\mu$m and 1cm and $\tau$=1 happens when $\Sigma_d$ $\approx$ 0.05 $\mathrm{g\, cm^{-2}}$ for \amax = 100$\mu$m and 1mm. This is corroborated in Figure \ref{fig:1pop} as the dips are very shallow when $\Sigma_{peak}$ = 0.02 $\mathrm{g\, cm^{-2}}$ and become much deeper when $\Sigma_{peak}$ = 0.2 $\mathrm{g\, cm^{-2}}$.

\begin{deluxetable*}{cccccc}
\tabletypesize{\scriptsize}
\tablecaption{Model list \label{table}}
\tablehead{
\colhead{Parameters} &
\colhead{$\Sigma_{peak}$ [$\mathrm{g\, cm^{-2}}$]} & \colhead{$\sigma$ [au]} & \colhead{$\Sigma_{floor}/\Sigma_{peak}$} & 
\colhead{$a_{max}$} &
\colhead{$\alpha$}
}
\colnumbers
\startdata
 one population & 0.02, 0.2, 2 & 5, 10, 20  & 10$^{-1}$, 10$^{-2}$, 10$^{-3}$, 10$^{-4}$ & 1, 10, 100 $\mu$m, 1 mm, 1 cm & --\\
 two population & 0.0002, 0.002, 0.02, 0.2& 5, 10, 20  & 10$^{-1}$, 10$^{-2}$, 10$^{-3}$, 10$^{-4}$ & \change{[0.1 $\mu$m to 1 $\mu$m]} and [100 $\mu$m to 1 cm] & 10$^{-2}$, 10$^{-3}$, 10$^{-4}$\\
\enddata
\end{deluxetable*}

\subsection{Two-population models \label{sec:twopop}}
In two-population models, we add a second species, with different opacity and density distributions. The first population is still assumed to be small grains and coupled with the gas. The second population is assumed to be large grains, partially decoupled with the gas, and concentrated more to the midplane and the ring center. Its vertical density distribution is a Gaussian with the scale height determined by the midplane temperature calculated from MCRT and the coupling parameter $\psi$ (\citealt{dullemond18b}, also in \citealt{zhu12}). 
The coupling parameter $\psi$ also determines the ring's width of the second species, and
\begin{equation}
    \psi = \sqrt{\frac{\alpha}{St}},
\end{equation}
where $\alpha$ is the disk turbulence viscosity, and $St$ is the Stokes number (or particle's dimensionless stopping time),
\begin{eqnarray}
St&=&t_{stop}\Omega=\frac{\pi a \rho_{p}}{2 \Sigma_{gas}} \nonumber\\
&=&1.57\times10^{-3}\frac{\rho_{p}}{1 \mathrm{g \,cm^{-3}}}\frac{a}{1 \mathrm{mm}}\frac{100 \mathrm{g\, cm^{-2}}}{\Sigma_{g}}, \label{eq:stokes}
\end{eqnarray}
where $\rho_p$ is the density of the dust particle, $a$ is the radius of the dust particle, and $\Sigma_{g}$ is the gas surface density. We adopt $\rho_p$=1.675 g cm$^{-3}$ as in the DSHARP opacity, and use $a$ = 1 mm to represent big grains. 
Effectively, $\psi$ is determined by the gas surface density $\Sigma_g$, the grain size $a$, and the disk viscosity $\alpha$. To explore the parameter space, we vary $\alpha$ and $\Sigma_g$ in the models.
The width and scale height of big grains are \citep{dullemond18b},
\begin{equation}
    \sigma_d = \sigma (1 + \psi^{-2})^{-1/2},
\end{equation}
and
\begin{equation}
    h_d = h_g (1 + \psi^{-2})^{-1/2}.
\end{equation}

The Gaussian peak is still centered at 70 au. At the Gaussian peak, we assume that the total dust to gas mass ratio $\epsilon$ = 1:100. At the Gaussian peak, the small and big grains' mass ratio is 1/31.6. We assume that the small grains have the same distribution as the gas. Since the distributions of big and small grains have different widths and scale heights, the local $\epsilon$ and the mass ratio between big and small grains vary at the 2-D r-$\theta$ plane.  Even though big grains contribute more mass inside the ring, the region outside the ring's midplane is still dominated by small grains, since small grains have a larger width and scale height.

For the two-population models, the second dust population is also involved in the iteration process in searching for the self-consistent thermal structure. In each iteration, besides adjusting the small grain and gas density, the big grains' scale height is changed due to the updated midplane temperature. Then the big dust density is adjusted vertically, but the surface density is always fixed. The convergence can also be reached after several iterations.

Our choices of $\alpha$ are 10$^{-2}$, 10$^{-3}$, and 10$^{-4}$. With weaker turbulence, the ring is radially narrower and vertically more settled. The peak surface density for small grains are  $\Sigma_{peak}$ = 0.0002, 0.002, 0.02 and 0.2  $\mathrm{g\, cm^{-2}}$. The peak surface density of the big grain population is 31.6 times higher. The total integrated dust mass of the big grain population is only around 10 times higher, since the small grain have wider radial distribution. The choices of widths and  $\Sigma_{floor}/\Sigma_{peak}$ (the density floor ratio applies to both small and big grains) are the same as one-population models. The opacities for big and small grains are also fixed, as shown in bottom panel of Figure \ref{fig:opacity_all}. These models are summarized in Table \ref{table}.

\begin{figure*}[t!]
\includegraphics[width=\linewidth]{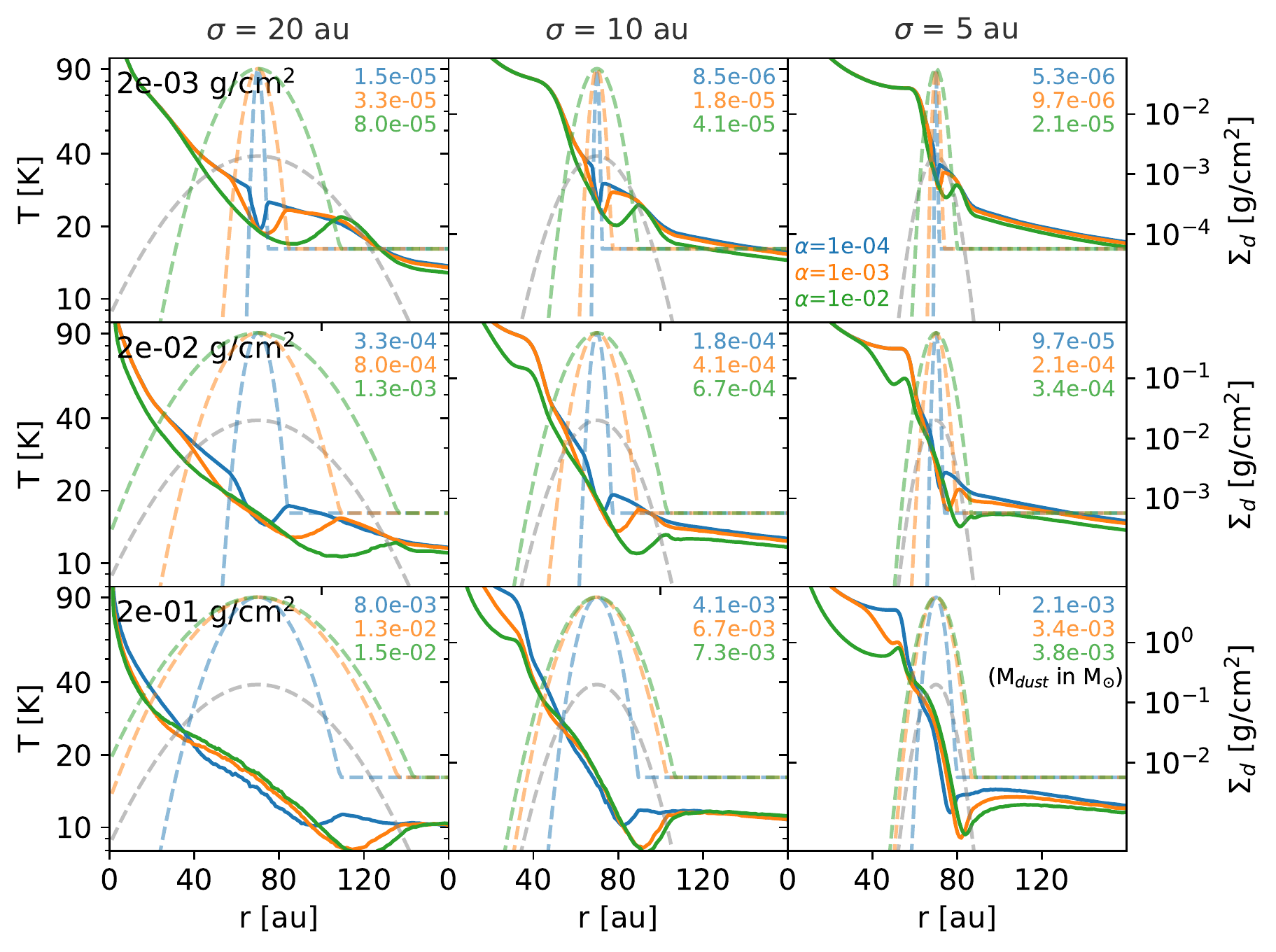}  
\figcaption{The midplane temperature of two population models with ring configurations. From top to bottom, the dust surface density for the small grains at the Gaussian peak are 0.002, 0.02 and 0.2 $\mathrm{g\, cm^{-2}}$ (the peak density of big grains are 31.6 times higher). From left to right, the Gaussian widths are 20, 10 and 5 au. Blue, orange and green curves represent $\alpha$=$10^{-4}$, $10^{-3}$, $10^{-2}$. The gray dashed lines are surface density of small grains and colored dashed lines are those of big grains. The floor density over the peak density is $10^{-3}$. The total dust masses are shown in the top right corner. In each panel, the dust mass depends on $\alpha$. The total masses are comparable to what are shown in each panel of Figure \ref{fig:1pop}.
\label{fig:2pop}}
\end{figure*}

Figure \ref{fig:2pop} shows the midplane temperatures for $\Sigma_{floor}/\Sigma_{peak}$ = 0.001 cases after iterations. The layout is the same as Figure \ref{fig:1pop}. From top to bottom, the peak surface density for the small grains are 0.002, 0.02 and 0.2 $\mathrm{g\, cm^{-2}}$. The peak densities of the big grains are larger by a factor of 31.6. The total dust masses are marked on the top-right corner of each panel and comparable to the respective panels in Figure \ref{fig:1pop}. Blue, orange and green curves represent $\alpha=$ $10^{-4}$, $10^{-3}$, $10^{-2}$ cases. The surface densities of the small grains are marked in gray dashed lines, and those of the big grains are marked in colored dashed lines. If the optical depth is low (the surface density is low and the ring is narrow), the temperature has a dip exactly located at the ring's position. The temperature at the dip can be 30\% lower than a smooth profile for the upper left cases. With a higher surface density and a wider ring, the dip moves towards the outer ring. When the surface density is high, the Stokes number becomes small and the big grains are more coupled to the gas. For the bottom panels, the temperature profiles are very similar to the one-population models (Figure \ref{fig:1pop}). The temperature dips occur outside the ring.
For reference, the Rosseland mean absorption opacity for the second species $k_{R, abs}$ = 8.13 $\mathrm{cm^{2}\, g^{-1}}$ at T=20 K. $\tau$ reaches unity when $\Sigma_{peak}$ for the small-grain population is around 0.005 $\mathrm{g\, cm^{-2}}$ (with total surface density reaching 0.15 $\mathrm{g\, cm^{-2}}$). This explains why the shadowing effect dominates when $\Sigma_{peak}$ $\gtrsim$ 0.02 $\mathrm{g\, cm^{-2}}$.
In short, the underlying mechanism that affects the midplane temperature becomes the shadowing effect in the optically thick regime or the big grains are well-coupled to the small grains. A main difference between two mechanisms is the position of the temperature dip. If the excess cooling dominates, the dip is at the ring's peak, whereas if the shadowing effect dominates, the dip is far outside the ring's peak. 

\begin{figure*}[t!]
\includegraphics[width=\linewidth]{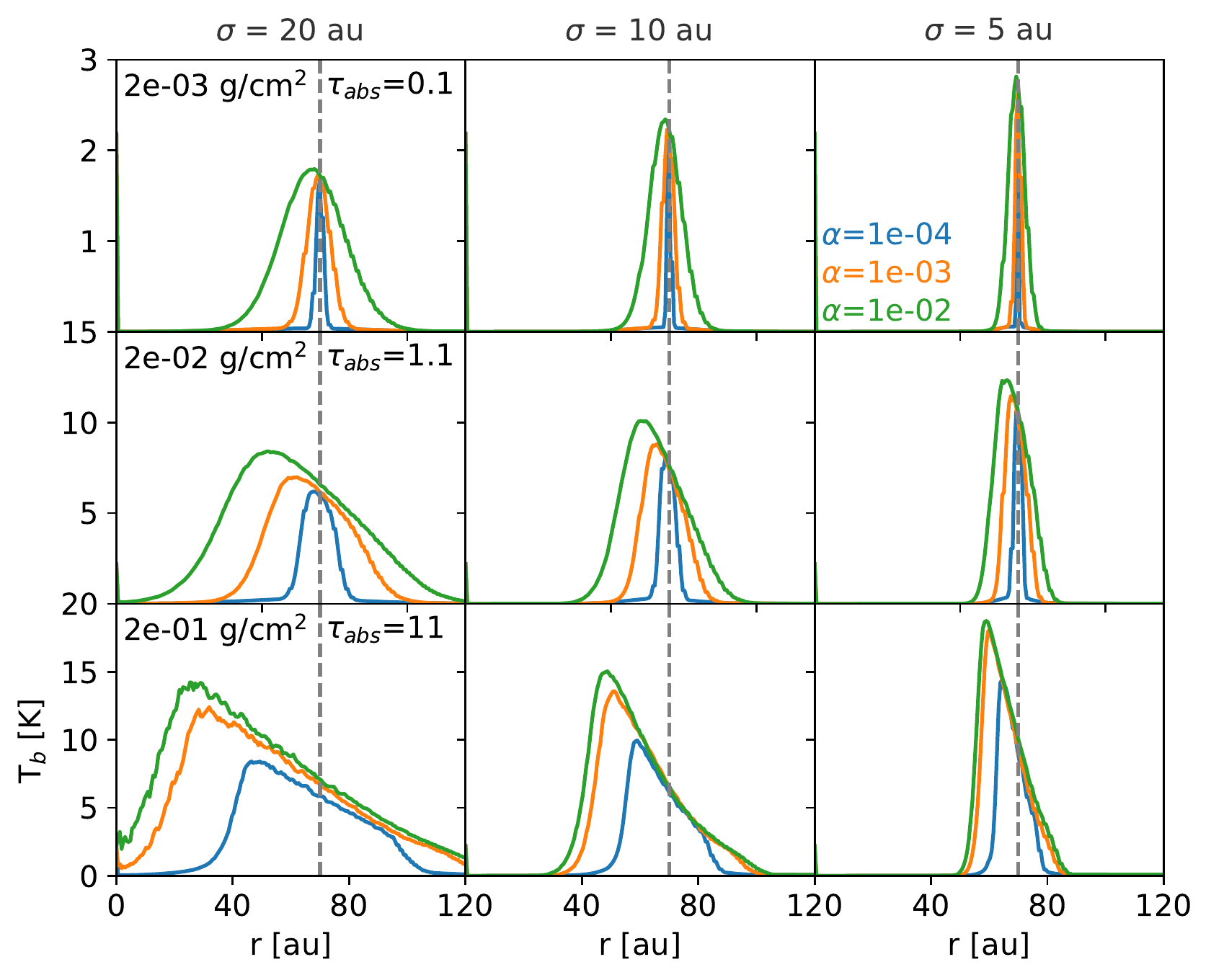}  
\figcaption{The radial profiles at 0.87 mm for two-population models in unit of brightness temperature as the same layout as Figure \ref{fig:2pop}. Total absorption optical depths of Guassian's peaks (including both species) are marked on left panels. The ring's peaks at 70 au are marked as vertical dashed lines.
\label{fig:2pop_intens}}
\end{figure*}

Figure \ref{fig:2pop_intens} shows the brightness temperature profiles for these models. The most prominent trend is that as the optical depth becomes higher, the ring's center shifts to the left. This is because when the ring becomes more optically thick, the temperature instead of the density profile is dominant in determining the radial profile. The higher temperature in the inner disk makes the peak shift inwards. The absorption optical depths at the peak of rings are marked on left panels. As expected, the transition happens at $\Sigma_{peak}$ = 0.02 $\mathrm{g\, cm^{-2}}$, as $\tau_{abs}$ is around unity. It is also possible that the temperature dip at the peak's center reduces the intensity and split a single ring into ring-gap-ring shape, but we do not observe it in these models. A simple test shows that the temperature dip needs to be very deep (more than 50\% decrease of temperature) to make it happen. The temperature decrease is at most 30\% among these models.

\subsection{2D thermal Structure}
\begin{figure*}[t!]
\includegraphics[width=\linewidth]{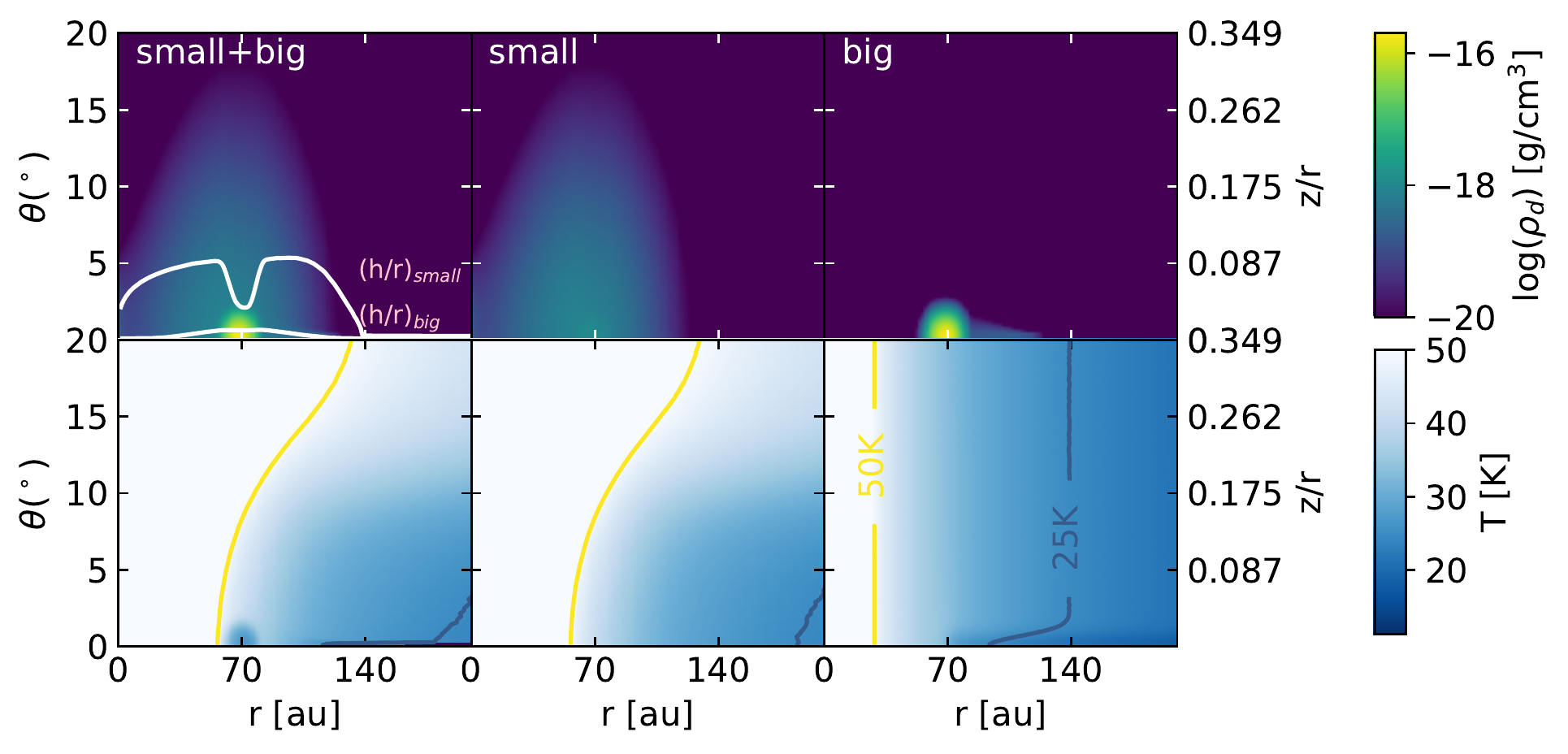}  
\figcaption{Left: 2D density (top) and temperature (bottom) distributions of model $\Sigma_{peak}$ = 2$\times 10^{-4}$ $\mathrm{g\, cm^{-2}}$, $\alpha$=0.01, $\sigma$ = 20 au and $\Sigma_{floor}/\Sigma{peak}$ = 0.001. Middle: the density and temperature distributions with only small grains included. Right: with only big grains included. The white curves in top panels are the one scale-height contours of small (upper one) and big (lower one) grains.
\label{fig:2d}}
\end{figure*}

Besides the midplane temperature, we also study the vertical temperature structure, which can be probed by molecular line observations (e.g.,\citealt{pinte18}).  In Figure \ref{fig:2d}, we plot the $r- \theta$ distributions of a model in the optically thin limit. The peak surface density for the small grain is $\Sigma_{peak}$ = 2$\times 10^{-4}$ $\mathrm{g\, cm^{-2}}$ (that of big grains is 31.6 times higher), $\alpha$=0.01, $\sigma$ = 20 au and $\Sigma_{floor}/\Sigma_{peak}$ = 0.001. The density and temperature maps are shown in the left panels. They have been iterated to reach hydrostatic equilibrium. Big grains are concentrated at 70 au with significant settling. The temperature is lower at the ring. The lower-temperature region extends vertically to $\sim$ 2$^\circ$ ($z/r$ $\sim$0.034), which is comparable to the local gas aspect ratio h/r. The positions of one scale height, $z$ = $h(r)$ are marked as white curves. Note that for small grains, the vertical density distribution is not a Gaussian, since the temperature varies along the disk height. The scale height is then defined as $z$ where
\begin{equation}
     \bigg[-2 \mathrm{ln}\Big(\frac{\rho(r, z)}{\rho(r, 0)} \Big) \bigg]^{1/2} = 1.
\end{equation}
The big grains' scale height is directly calculated from the midplane temperature and the coupling parameter $\psi$. 
At the ring, the scale heights of both species become lower due to the decrease of temperature there. 
For comparison, the upper-middle and upper-right panels show the density structures of the small-grain and big-grain components of the model, respectively. 
The lower-middle and lower-right panels are the temperatures of the disk if there is only small-grain or big-grain population. Note that their thermal structures are not self-consistently calculated, i.e., their temperatures are calculated using RADMC-3D without iterations and the density vertical structures are unchanged compared to the two-population case.
Since the small-grain-only model's temperature structure is almost identical to the mixed model containing both species, it is clear that almost all the disk temperature structure is determined by the small grains. The exception is at the ring where big grains are concentrated. The temperature there is much lower. If the disk only has big grains, the radial temperature profiles at different scale heights are similar (except at the ring), i.e., the vertical temperature distribution is close to be isothermal.


\begin{figure}[t!]
\includegraphics[width=\linewidth]{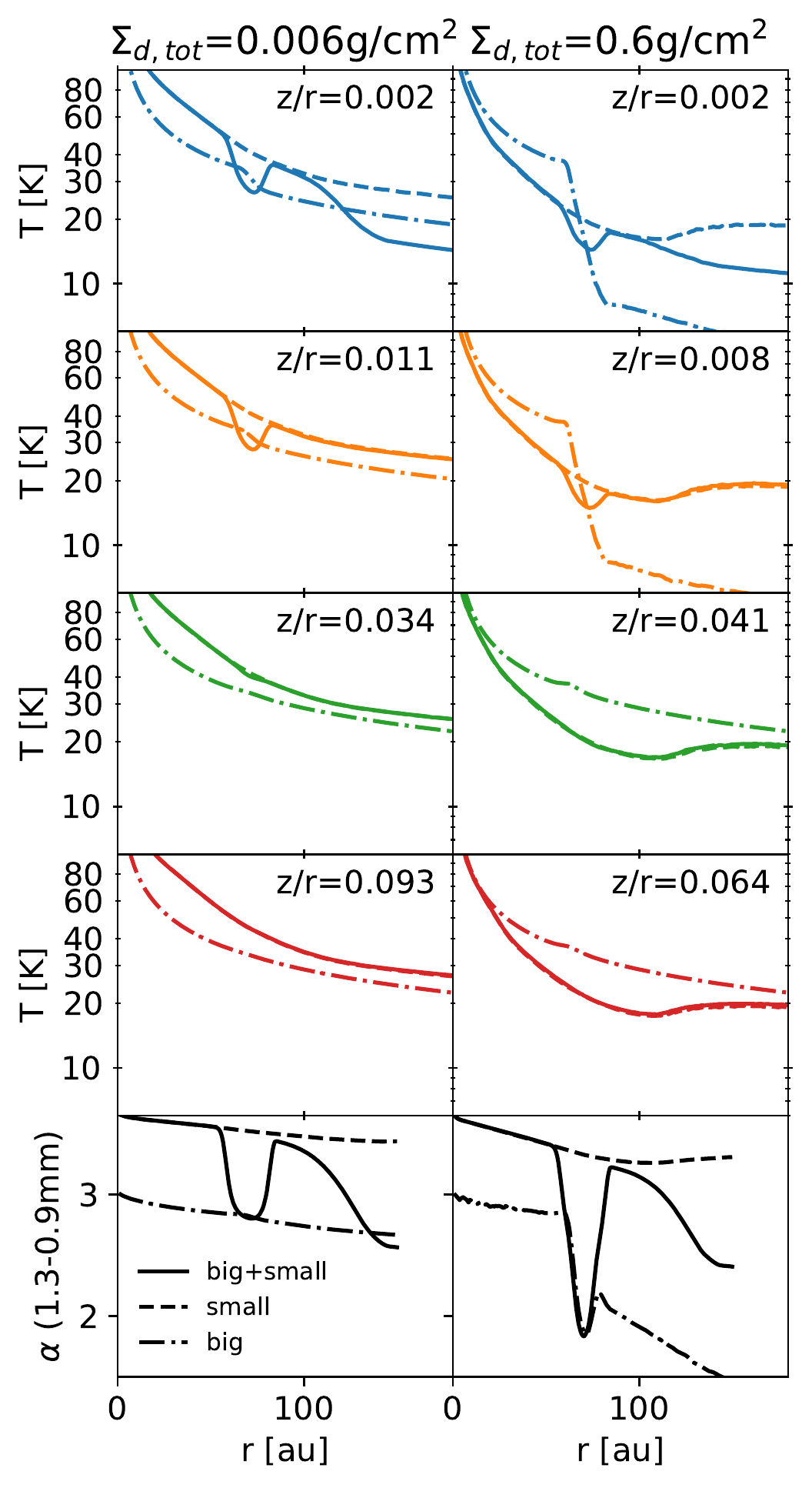}  
\figcaption{The radial temperature profiles at different heights for the setup of Figure \ref{fig:2d} (left) and disks with 100$\times$ higher density (right), and $\alpha=10^{-4}$. From top to bottom, the heights are picked at the disk's midplane, big grains' scale height, gaseous scale height and upper atmosphere. The bottom panels show the spectral index between 1.25 mm and 0.87 mm. Solid lines are the MCRT results with both big and small grains. Dashed lines are the temperatures profiles for MCRT with only small grains included. Dotted-dashed curves show big-grain-only cases.
\label{fig:radialdiffheight}}
\end{figure}

We also run a case with 100 times higher surface density. The turbulent viscosity $\alpha$ is $10^{-4}$ so that the big grains' width is comparable to the previous one. The ring becomes optically thick, and has a different behavior in the disk atmosphere.
Figure \ref{fig:radialdiffheight} shows the radial temperature profiles at different disk heights $\theta$ (or $z/r$) for these two models. The temperatures of low-mass disks are shown in the left panels and those of high-mass disks are shown in the right panels. The top four panels show the temperatures measured at the midplane, big grains' scale height, small grains' scale height (or gas scale height) and a location at a higher atmosphere. The temperature dips are obvious within the big grains' scale height for both low-mass and high-mass disks. This means that excess cooling is operating in both disks' midplane. At larger vertical heights, the temperature decreases smoothly with radius for the low-mass disk, whereas it has a dip outside the ring around 100 au for the high-mass disk. The latter is essentially a one-population scenario at these scale heights where small grains dominate. The small grains' ring is optically thick enough for the shadowing effect to operate. For reference, the small-grain-only and big-grain-only temperatures using MCRT are marked as dashed and dotted-dashed curves. In the low-mass disk case, the small-grain-only temperature is higher than the big-grain-only temperature, while it is the opposite in the high-mass disk case.  The bottom panels show the spectral indices for these models between ALMA band 6 and 7. Both models see the drop of spectral indices within the ring. The dip in the spectral index of the low-mass disk indicates that the ring is dominated by big grains, whereas the rest is dominated by small grains. For the high-mass disk, the spectral index $\alpha$ is even lower than two at the ring, which indicates that the ring is optically thick and dust scattering is substantial \citep{zhu19b,Liu2019}. For these two configurations, the temperature dip due to excess cooling cannot be seen at one gas scale height for the two-population scenario. If the disk is optically thick, the temperature dip outside the ring can still be seen high above the midplane. At the big grain's scale height (z/r=0.008), the temperature profile even has two dips. One is at the ring position due to excess cooling from big grains, whereas the other is outside the ring due to the shadowing effect.

\section{Disk Thermal structure coupled with a dust evolutionary model \label{sec:dustevomodel}}
In both optically thin and thick cases, the disk temperature drops either at the ring or at the outer edge of the ring. In either cases, the temperature dip acts as a pressure trap, which can alter the shape of the ring. In this section we attempt to add dust growth and evolution in the feedback loop to test this scenario. The detail of the 1D dust evolution code is summarized in Appendix \ref{sec:1ddustevo}.

The gas surface density profile is set to resemble that of a transition disk, LkCa 15 \citep{facchini20} with a depleted inner cavity,
\begin{equation}
\Sigma_g(r) = \Sigma_{g,c} \Big(\frac{r}{65 \mathrm{au}}\Big)^{p} \mathrm{arctan}\Big[\Big(\frac{r}{65 \mathrm{au}}\Big)^{10}\Big], 
\end{equation}
where $p$ is the slope of the surface density profile. We run dust evolution with a fixed temperature profile (T$\propto r^{-0.5}$) for 0.6 Myr. The gas surface density is fixed over time.
At t=0.6 Myr, the dust has already piled up at the outer edge of the inner cavity and formed a Gaussian ring. The grain size has grown to $\sim$ mm size at the ring. The evolving dust distribution should change the temperature profile, which in turn, alters the dust distribution. Starting from this point, the temperature is calculated using MCRT self-consistently as described in Section \ref{sec:twopop}. The dust surface density in the dust evolution code is used as the big-grain population (with \{\amin, \amax\}=\{0.1mm, 10mm\}). When the coupling parameter $\psi$ is involved, we assume all the big grains are 1 mm in the MCRT calculations. The small grain's density has the same profile as that of the gas. Their opacities are the same as mentioned in Section \ref{sec:twopop} and bottom of Figure \ref{fig:opacity_all}. The surface density ratio between two populations is still calculated assuming power-law dust size distribution with a slope of 3.5 at the ring's peak. The derived midplane temperature is then used in the dust evolution code. The temperature is updated using MCRT every time interval $dt_{MCRT}$. 

The convergence requires very high spatial and temporal resolutions. To resolve the dust evolution around the ring, we have 1024 radial bins uniformly spaced in $log(r)$ from 40 au to 300 au to make sure that each newly formed ring (if any) is resolved by at least 10 gird cells. The $dt_{MCRT}$ needs to be small enough to capture particles' radial drift. \change{The timescale for the disk to reach thermal equilibrium varies in several orders of magnitude across the disk. \citet{bae21} calculates the thermal relaxation timescale consisting of radiation, diffusion and gas-dust collisions. In their setup, the relaxation timescale at the midplane is comparable to the orbital timescale at 70 au, which is around 600 years. The particle's drift timescale is usually less than the orbital timescale (i.e., $St \lesssim$ 1). } The 2D MCRT calculation is much more numerically expensive than the 1D dust evolution.

\begin{figure*}[t!]
\includegraphics[width=\linewidth]{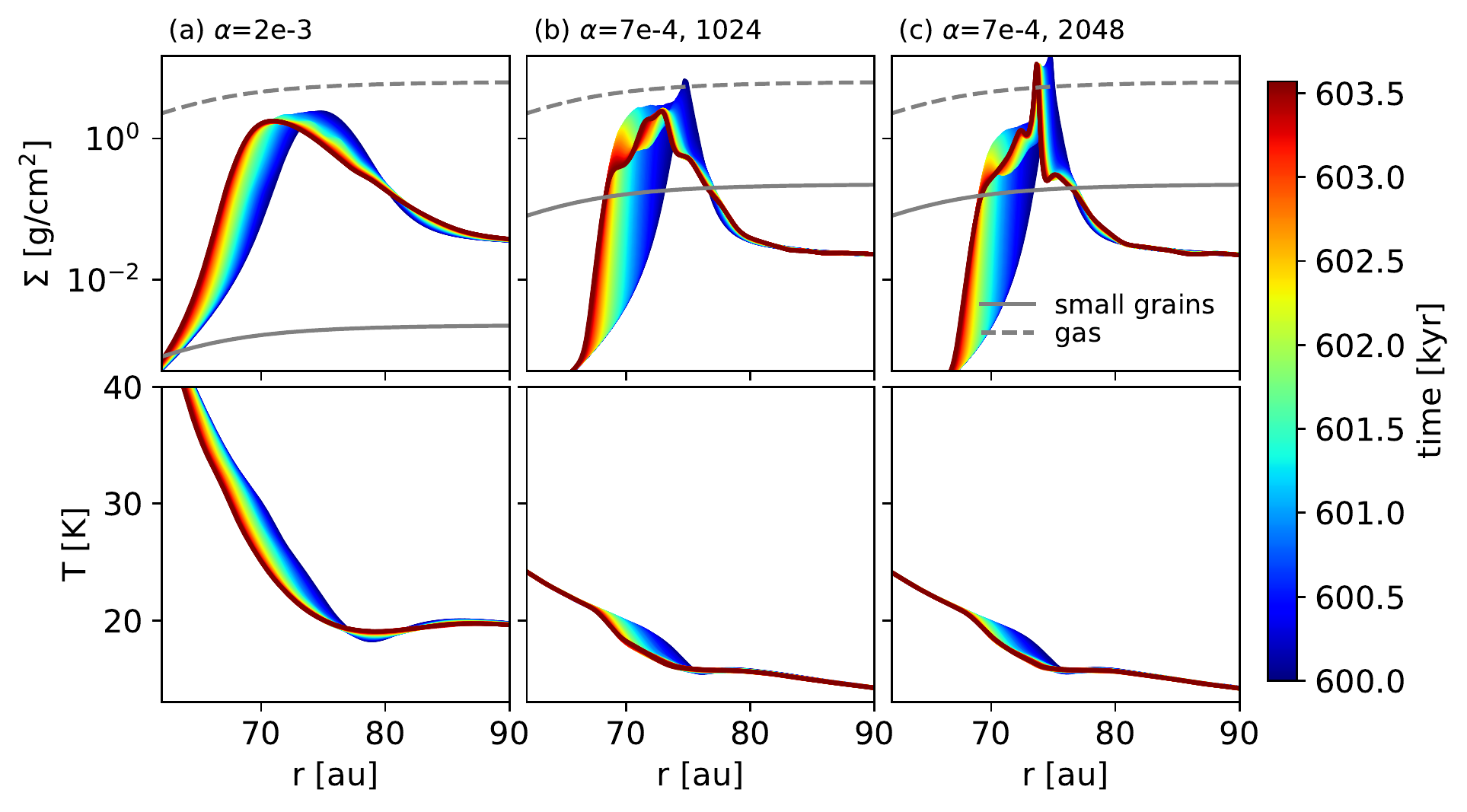}  
\figcaption{Dust surface density of the big-grain population (top panels) and midplane temperature (bottom panels). (a) Model with $\alpha$=2$\times 10^{-3}$, $\Sigma_{g}$=4.1 $\mathrm{g\, cm^{-2}}$, $\epsilon$ = 0.01 at the ring. (b, c) models with $\alpha$=7$\times 10^{-4}$ and with different radial resolutions. In these models small grains' surface density is 100$\times$ higher than that in model (a). The small grains' densities are marked by gray solid curves and gas surface densities are marked by gray dashed curves.
\label{fig:twomodels}}
\end{figure*}

In Figure \ref{fig:twomodels}, we present two models that evolve with $dt_{MCRT}$ = 30 yr for 9 kyr starting from t=0.6 Myr. Both models have gas surface density $\Sigma_{g,c}$ = 4.1 $\mathrm{g\, cm^{-2}}$, and $p$=0, so that the pressure gradient around the ring solely comes from the temperature profile. The surface density of big grains is represented by colored curves, whereas small grain and gas components are represented by solid and dashed gray curves. Model (a) has $\alpha$=2$\times 10^{-3}$. In this case, the turbulence is too strong to split a ring into more rings. Nevertheless, the initial Gaussian ring tilts towards the inner disk as the disk evolves. This skewed shape that deviates from a  Gaussian profile is seen in HD 163296 B67 ring \citep{dullemond18b, isella18b}. Note that, even without considering the temperature feedback effect discussed here, any non-Gaussian shape of the pressure bump  can also lead to this skewed profile of dust distribution. This skewed dust shape can also occur when the dust is drifting to the ring center before reaching the steady state.  At the initial stage, the temperature dip is at the outer edge of the ring. This means that the mechanism for the temperature dip is due to the shadowing effect. This is not surprising since big grains always dominate in the area of interest. At a later stage, the temperature dip becomes shallower and smoother, which indicates a negative feedback. This negative feedback leads to a steady state where the radial pressure gradient gradually becomes zero.

To come up with a condition that excess cooling can operate in producing the temperature dip, we increase the small grains' surface density by a factor of 100 in model (b). Now big grains only dominate inside the ring, whereas small grains dominate outside the ring. In addition, we also lower the viscosity $\alpha$ to 7$\times 10^{-4}$, hoping to generate more substructures. After several thousand years, the initial ring evolves to several rings. With a larger optical depth, the temperature is thus much lower than the previous case. The initial temperature dip is close to the ring, which means that excess cooling is indeed operating. Due to the high computational cost of MCRT, the dust evolution is stopped after 3.5 kyr. At the end of simulations, neither of these two models reach steady states. We are expecting more dust will be trapped at the ring. On the other hand, this pile-up may last over the disk's lifetime and the disk may not reach any steady state eventually. To test the resolution effect, we increase the radial resolution to 2048 in panel (c). The substructure persists in the higher resolution run.

Finally, self-consistent ring structure model including both thermal effects and dust dynamics is important not only for protoplanetary disk observations but also for planetesimal and planet formation studies \citep{Morbidelli2020,chambers2021}. The formation of skewed rings and even multiple rings will affect the mass and number of formed planetesimals. Converting dust into planetesimals can also affect the dust opacity and feedback to the ring's thermal structure. More work on this feedback loop on planetesimal formation needs to be studied in future.

\section{Discussion \label{sec:discussion}}
\subsection{Impact of density floor}
\begin{figure*}[t!]
\includegraphics[width=\linewidth]{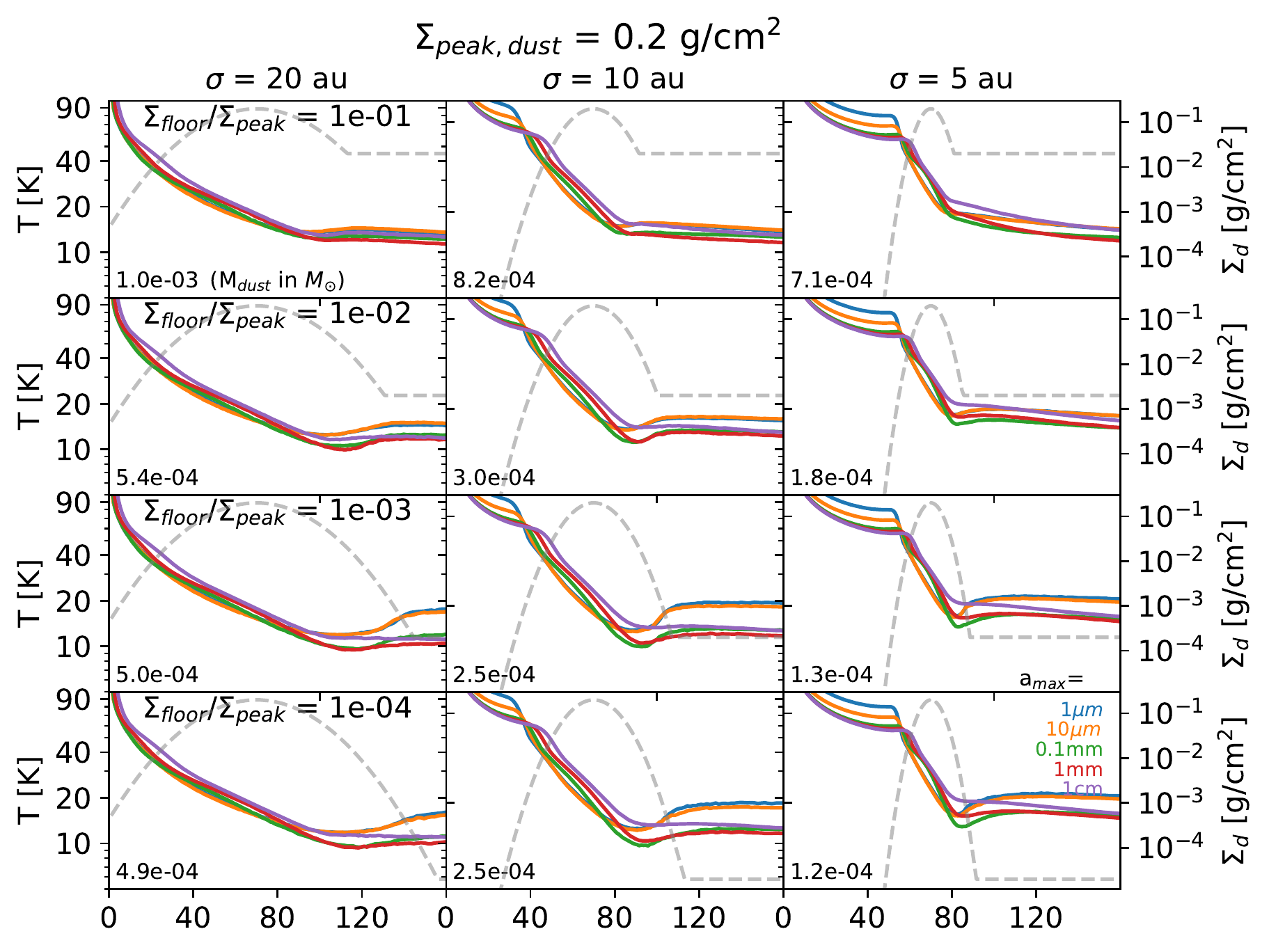}
\figcaption{The effects of density floor on temperature profiles for one population. The dust peak surface density is 0.2 $\mathrm{g\, cm^{-2}}$. From top to bottom, the density floors are 0.1, 0.01, 0.001 and 0001. The temperature dip becomes stronger with lower density floor.
\label{fig:floor}}
\end{figure*}

We have studied how the density outside the Gaussian peak can affect the disk temperature.
We assume that the dust density  levels off at some distance away from the peak. We take $\Sigma_{floor}/\Sigma_{peak}$ = $10^{-3}$ as our fiducial model. This value affects the temperature gradient for the shadowing effect. We present other $\Sigma_{floor}/\Sigma_{peak}$ values for single population runs with $\Sigma_{peak}$ = 0.2 $\mathrm{g\, cm^{-2}}$ in Figure \ref{fig:floor}. From top to bottom, the floor floor becomes lower. The temperature gradient becomes larger with lower density floor, but the change becomes insignificant when $\Sigma_{floor}/\Sigma_{peak}\le$  $10^{-3}$. When the floor is higher, the temperature dip is shallower due to a less drastic change of the density profile. A higher density floor makes the density gradient smoother. This implies that if there are dust grains being constantly replenished in the outer disk, the temperature dip would be shallow.

\subsection{Temperature gap in CI Tau}
During the preparation of this manuscript, a line emission gap in a continuum ring has been observed \citep{rosotti21} in $^{13}$CO emission of CI Tau disk. CI Tau is a disk that has been observed with substructures in dust continuum \citep{clarke18}. There are at least three rings at 23, 54 and 135 au. This temperature gap is located at the second continuum ring at 54 au \citep{rosotti21}. The authors exclude the possibility of reduction in surface density and explore the possibility of reduction in temperature. To reproduce a temperature gap, they multiply the small dust surface density at 20 au, with a Gaussian width $\sigma$ = 10 au. It casts a shadow onto the outer disk, which has a lower temperature and is positioned around 54 au and thus explains the temperature gap. Their explanation falls into our one-population scenario. If this is the case, the location of this temperature gap matches the second ring purely by chance. On the other hand, we propose that our two-population scenario can naturally explain why the temperature gap is exactly at the position of the ring. As mentioned in previous sections, big grains dominate in the ring and the cooling is more efficient. Thus, the temperature in the ring is lower. If this mechanism dominates, we predict that the temperature gap will disappear if one uses a more optically thick tracer to probe higher emission surfaces. On the other hand, if this is indeed due to the shadowing effect suggested by \cite{rosotti21}, the temperature gap should still be able to be observed at several disk scale heights.

With more high-resolution line emission data available in future, we may be able to find more cases showing temperature gaps at or beyond a continuum ring, since both the shadowing effect and excess cooling can lead to a temperature gap.

\section{Conclusion \label{sec:conclusion}}
We study the thermal structure of a dusty ring self-consistently by iterating the disk's dust distribution, thermal structure from MCRT, and vertical hydrostatic equilibrium. We find that two different effects can lead to a temperature dip in a ring. One is the shadowing effect and the other is excess cooling of big grains concentrated in rings. The latter effect is studied for the first time in this work.

We include two dust populations in our model: small and big grains. For each population, we adopt a Gaussian ring structure in the radial direction. Due to the dust trapping by the pressure bump, the small and big grains have different radial widths and scale heights, depending on their coupling to the gas.  After several iterations, we find that temperature drops substantially inside the ring, which is due to the different equilibrium temperatures of small and big grains. With the higher opacity at submm/mm wavelengths, big grains have higher cooling efficiency, so that the temperature drops at the ring center where they are concentrated.  

The temperature dip by the shadowing effect is important when 
the disk is optically thick. Its temperature dip exists in a large vertical region. On the other hand, the temperature dip by excess cooling of big grains is the strongest when the disk is optically thin, and it happens closer to the midplane. In reality, both mechanisms operate together to some degree. Both mechanisms can explain the temperature gap observed in CI Tau \citep{rosotti21}, but the excess cooling effect does not need fine-tuning. The gap is located at the dust continuum ring, which is the exact prediction of this excess cooling mechanism, instead of a coincidence as in the shadowing mechanism.

Temperature dips by either of these two mechanisms can possibly lead to the formation of more rings beyond the initial dust ring. This is due to the feedback from the disk's thermal structure to the dust radial drift. The temperature bump can produce a pressure bump, given that the gas surface density does not vary abruptly. We demonstrate this possibility by combining the MCRT calculation with a 1D dust evolution code. If the drift timescale is long, the ring will not split into more rings. Instead, it will deviate from a Gaussian, with steeper inner edge and shallower outer edge.  With a low viscosity and different dominant populations inside and outside of a ring, one ring can indeed evolve to more rings. The separation is around several au, which is within the observational limit of ALMA. We have already seen that with higher resolutions that previously observed wide single ring can be resolved into more rings (e.g., \citealt{facchini20, benisty21}). More systems with these closely packed rings may be found in future higher resolution observations. By constructing self-consistent ring models and comparing with observations, we may be able to constrain the dust size distribution, dust settling, and dust radial drift in rings of protoplanetary disks. These information can be used to constrain planetesimal and planet formation within rings.

\acknowledgments We thank insightful comments from an anonymous referee. S. Z. thanks helpful discussion with Lee Hartmann, Ruobing Dong, Xue-Ning Bai and Chris Ormel. S. Z. acknowledges supports from University of Nevada, Las Vegas Barrick Fellowship and by NASA through the NASA FINNEST grant 80NSSC20K1376. X. H. thanks helpful discussion with Satoshi Okuzumi on dust modelling. X. H. acknowledges support by NSF AST-1716259 and the University of Virginia through a Virginia Initiatives on Cosmic Origins grant. Z. Z. acknowledges support from the National Science Foundation under CAREER grant AST-1753168. J. B. acknowledges support by NASA through the NASA Hubble Fellowship grant \#HST-HF2-51427.001-A awarded  by  the  Space  Telescope  Science  Institute,  which  is  operated  by  the  Association  of  Universities  for  Research  in  Astronomy, Incorporated, under NASA contract NAS5-26555. 
All simulations are carried out using computers supported by the
Texas Advanced Computing Center (TACC) at the University of
Texas at Austin through XSEDE grant TG-AST130002, TG-AST200032 and from
the NASA High-End Computing (HEC) program through the NASA
Advanced Supercomputing (NAS) Division at Ames Research Center.   
\software{
{\tt RADMC-3D} \citep{dullemond12},
{\tt dsharp\_opac} \citep{github_dsharp_opac},
{\tt Matplotlib} \citep{matplotlib},
{\tt Numpy} \citep{numpy}, 
{\tt Scipy} \citep{scipy}
}

\appendix
\section{1D dust evolution model \label{sec:1ddustevo}}

 We use a single size approximation for dust evolution calculation, similar to \citet{2016A&A...589A..15S}. One of the motivations of this approach is that the mass-size distribution of dust is top heavy, so the dust surface density can be well represented by its maximum sized species. In cylindrical coordinates, the advection-diffusion equation for dust surface density $\Sigma_d$ is
\begin{equation}
\frac{\partial \Sigma_d}{\partial t}+\frac{1}{r}\frac{\partial}{\partial r}\left[rv_r\Sigma_d-\frac{\nu}{1+{\rm St}^2}r\Sigma_g\frac{\partial Z}{\partial r}\right]=0,
\label{eq:sigma_d_adv}
\end{equation}
where $v_r$ is grain's radial velocity, $\nu$ is the
turbulent viscosity, ${\rm St}$ is the dimensionless stopping time of the
dust particles, and $Z$ is the dust-to-gas surface density ratio. The exact value of $v_r$ is governed by angular momentum loss by gas head wind \citep{1977MNRAS.180...57W}.
\begin{equation}
v_r=\frac{1}{{\rm St}+{\rm St}^{-1}}\eta v_K;\  {\rm St}=\frac{\pi}{2}\frac{\rho_da}{\Sigma_g}\times {\rm max}\left[ 1, \frac{4a}{9\lambda}\right],
\end{equation}
where $v_K$ is the local Keplerian orbital velocity, $\rho_d$ is mass density of individual particle, $a$ is the dust particle radius, and $\lambda$ is the mean free path of the gas. Stokes number St covers two drag regimes: Reynolds regime and the first Stokes regime. $\eta$ is the parameter that reflects the gas disk's pressure gradient,
\begin{equation}
\eta=\frac{d\ln{(c_s^2\rho_g)}}{d\ln{r}}\left(\frac{c_s}{v_K}\right)^2.
\end{equation}

The core of the single sized approximation is the equation governing the size evolution of a representing particle,
\begin{equation}
\frac{\partial m_p}{\partial t}+v_r\frac{\partial m_p}{\partial r}=\Sigma_d\frac{2\sqrt{\pi}a_p^2v_{pp}}{h_d}.
\end{equation}
The source term on the right hand side is the growth kernel that defines how fast this particle sweeps up mass. The particle-particle velocity, $v_{pp}$ can be divided into five components,
\begin{equation}
    v_{pp}=\sqrt{(\Delta v_B)^2+(\Delta v_t)^2+(\Delta v_r)^2+(\Delta v_\phi)^2+(\Delta v_z)^2}.
\end{equation}
The first term is mutual velocity from Brownian motion. If the mass of the other particle is $m_p'$, then $\Delta v_B=\sqrt{8(m_p+m_p')k_BT/(\pi m_pm_p')}$. Since $\Delta v_B$ does not diminish to zero when $m_p=m_p'$, we use $\Delta v_B=4\sqrt{k_BT/(\pi m_p)}$ with $m_p$ as the mass of the representing size in our single-sized model. 
The second term is velocity from turbulent mixing. For two particles with Stokes number ${\rm St}$ and ${\rm St}'$ in a disk with $\alpha$ viscosity,
\begin{equation}
\Delta v_t=
\begin{cases}
\sqrt{\alpha}c_s{\rm Re_t}^{1/4}({\rm St}-{\rm St}')&\text{if }  {\rm St}\ll 1/\sqrt{Re_t}, \\
(1.4...1.7)\times \sqrt{\alpha}c_s \sqrt{\rm St} & \text{if } 1/\sqrt{\rm Re_t}\ll {\rm St} \ll 1,\\
\sqrt{\alpha}c_s\sqrt{\frac{1}{1+{\rm St}}+\frac{1}{1+{\rm St}'}} & \text{if } {\rm St} \gg 1,
\end{cases}
\end{equation}
where ${\rm Re_t} = 2\nu/ v_{th}\lambda$ is the turbulent Reynolds number, and $v_{th} =\sqrt{8k_BT/\pi\mu}$ is the thermal velocity of the gas. Note that the first function diminishes to zero when ${\rm St}={\rm St}'$. This is the same for $\Delta v_r, \Delta v_\phi, \Delta v_z$. When this happens one needs to use a second representing particle size to calculate ${\rm St}'$, but does not need to evolve the second particle independently. Following \citet{2016A&A...589A..15S}, this single sized model produces the best result when ${\rm St}'/{\rm St}=0.5$ comparing with a full dust evolution model.

\clearpage


\begin{thebibliography}{}
\expandafter\ifx\csname natexlab\endcsname\relax\def\natexlab#1{#1}\fi
\providecommand{\url}[1]{\href{#1}{#1}}
\providecommand{\dodoi}[1]{doi:~\href{http://doi.org/#1}{\nolinkurl{#1}}}
\providecommand{\doeprint}[1]{\href{http://ascl.net/#1}{\nolinkurl{http://ascl.net/#1}}}
\providecommand{\doarXiv}[1]{\href{https://arxiv.org/abs/#1}{\nolinkurl{https://arxiv.org/abs/#1}}}

\bibitem[{{ALMA Partnership} {et~al.}(2015){ALMA Partnership}, {Brogan},
  {P{\'e}rez}, {Hunter}, {Dent}, {Hales}, {Hills}, {Corder}, {Fomalont},
  {Vlahakis}, {Asaki}, {Barkats}, {Hirota}, {Hodge}, {Impellizzeri}, {Kneissl},
  {Liuzzo}, {Lucas}, {Marcelino}, {Matsushita}, {Nakanishi}, {Phillips},
  {Richards}, {Toledo}, {Aladro}, {Broguiere}, {Cortes}, {Cortes}, {Espada},
  {Galarza}, {Garcia-Appadoo}, {Guzman-Ramirez}, {Humphreys}, {Jung}, {Kameno},
  {Laing}, {Leon}, {Marconi}, {Mignano}, {Nikolic}, {Nyman}, {Radiszcz},
  {Remijan}, {Rod{\'o}n}, {Sawada}, {Takahashi}, {Tilanus}, {Vila Vilaro},
  {Watson}, {Wiklind}, {Akiyama}, {Chapillon}, {de Gregorio-Monsalvo}, {Di
  Francesco}, {Gueth}, {Kawamura}, {Lee}, {Nguyen Luong}, {Mangum}, {Pietu},
  {Sanhueza}, {Saigo}, {Takakuwa}, {Ubach}, {van Kempen}, {Wootten},
  {Castro-Carrizo}, {Francke}, {Gallardo}, {Garcia}, {Gonzalez}, {Hill},
  {Kaminski}, {Kurono}, {Liu}, {Lopez}, {Morales}, {Plarre}, {Schieven},
  {Testi}, {Videla}, {Villard}, {Andreani}, {Hibbard}, \&
  {Tatematsu}}]{ALMA2015}
{ALMA Partnership}, {Brogan}, C.~L., {P{\'e}rez}, L.~M., {et~al.} 2015, \apjl,
  808, L3, \dodoi{10.1088/2041-8205/808/1/L3}

\bibitem[{Andrews(2020)}]{andrews20}
Andrews, S.~M. 2020, Annual Review of Astronomy and Astrophysics, 58, 483,
  \dodoi{10.1146/annurev-astro-031220-010302}

\bibitem[{{Andrews} {et~al.}(2012){Andrews}, {Wilner}, {Hughes}, {Qi},
  {Rosenfeld}, {{\"O}berg}, {Birnstiel}, {Espaillat}, {Cieza}, {Williams},
  {Lin}, \& {Ho}}]{andrews12}
{Andrews}, S.~M., {Wilner}, D.~J., {Hughes}, A.~M., {et~al.} 2012, \apj, 744,
  162, \dodoi{10.1088/0004-637X/744/2/162}

\bibitem[{{Andrews} {et~al.}(2018){Andrews}, {Huang}, {P{\'e}rez}, {Isella},
  {Dullemond}, {Kurtovic}, {Guzm{\'a}n}, {Carpenter}, {Wilner}, {Zhang}, {Zhu},
  {Birnstiel}, {Bai}, {Benisty}, {Hughes}, {{\"O}berg}, \&
  {Ricci}}]{andrews18b}
{Andrews}, S.~M., {Huang}, J., {P{\'e}rez}, L.~M., {et~al.} 2018, \apjl, 869,
  L41, \dodoi{10.3847/2041-8213/aaf741}

\bibitem[{{Bae} {et~al.}(2021){Bae}, {Teague}, \& {Zhu}}]{bae21}
{Bae}, J., {Teague}, R., \& {Zhu}, Z. 2021, \apj, 912, 56,
  \dodoi{10.3847/1538-4357/abe45e}

\bibitem[{{Bae} {et~al.}(2019){Bae}, {Zhu}, {Baruteau}, {Benisty}, {Dullemond},
  {Facchini}, {Isella}, {Keppler}, {P{\'e}rez}, \& {Teague}}]{bae19}
{Bae}, J., {Zhu}, Z., {Baruteau}, C., {et~al.} 2019, \apjl, 884, L41,
  \dodoi{10.3847/2041-8213/ab46b0}

\bibitem[{{Benisty} {et~al.}(2021){Benisty}, {Bae}, {Facchini}, {Keppler},
  {Teague}, {Isella}, {Kurtovic}, {P{\'e}rez}, {Sierra}, {Andrews},
  {Carpenter}, {Czekala}, {Dominik}, {Henning}, {Menard}, {Pinilla}, \&
  {Zurlo}}]{benisty21}
{Benisty}, M., {Bae}, J., {Facchini}, S., {et~al.} 2021, \apjl, 916, L2,
  \dodoi{10.3847/2041-8213/ac0f83}

\bibitem[{Birnstiel(2018)}]{github_dsharp_opac}
Birnstiel, T. 2018, {dsharp\_opac: The DSHARP Mie-Opacity Library}, 1.1.0,
  Zenodo, \dodoi{10.5281/zenodo.1495277}

\bibitem[{{Birnstiel} {et~al.}(2018){Birnstiel}, {Dullemond}, {Zhu}, {Andrews},
  {Bai}, {Wilner}, {Carpenter}, {Huang}, {Isella}, {Benisty}, {P{\'e}rez}, \&
  {Zhang}}]{birnstiel18}
{Birnstiel}, T., {Dullemond}, C.~P., {Zhu}, Z., {et~al.} 2018, \apjl, 869, L45,
  \dodoi{10.3847/2041-8213/aaf743}

\bibitem[{{Calvet} {et~al.}(1991){Calvet}, {Patino}, {Magris}, \&
  {D'Alessio}}]{calvet91}
{Calvet}, N., {Patino}, A., {Magris}, G.~C., \& {D'Alessio}, P. 1991, \apj,
  380, 617, \dodoi{10.1086/170618}

\bibitem[{{Carrasco-Gonz{\'a}lez} {et~al.}(2019){Carrasco-Gonz{\'a}lez},
  {Sierra}, {Flock}, {Zhu}, {Henning}, {Chandler}, {Galv{\'a}n-Madrid},
  {Mac{\'\i}as}, {Anglada}, {Linz}, {Osorio}, {Rodr{\'\i}guez}, {Testi},
  {Torrelles}, {P{\'e}rez}, \& {Liu}}]{carrasco2019}
{Carrasco-Gonz{\'a}lez}, C., {Sierra}, A., {Flock}, M., {et~al.} 2019, arXiv
  e-prints, arXiv:1908.07140.
\newblock \doarXiv{1908.07140}

\bibitem[{{Chambers}(2021)}]{chambers2021}
{Chambers}, J. 2021, \apj, 914, 102, \dodoi{10.3847/1538-4357/abfaa4}

\bibitem[{{Chiang} \& {Goldreich}(1997)}]{chiang97}
{Chiang}, E.~I., \& {Goldreich}, P. 1997, \apj, 490, 368,
  \dodoi{10.1086/304869}

\bibitem[{{Clarke} {et~al.}(2018){Clarke}, {Tazzari}, {Juhasz}, {Rosotti},
  {Booth}, {Facchini}, {Ilee}, {Johns-Krull}, {Kama}, {Meru}, \&
  {Prato}}]{clarke18}
{Clarke}, C.~J., {Tazzari}, M., {Juhasz}, A., {et~al.} 2018, \apjl, 866, L6,
  \dodoi{10.3847/2041-8213/aae36b}

\bibitem[{{Dong} {et~al.}(2018){Dong}, {Liu}, {Eisner}, {Andrews}, {Fung},
  {Zhu}, {Chiang}, {Hashimoto}, {Liu}, {Casassus}, {Esposito}, {Hasegawa},
  {Muto}, {Pavlyuchenkov}, {Wilner}, {Akiyama}, {Tamura}, \&
  {Wisniewski}}]{dong18a}
{Dong}, R., {Liu}, S.-y., {Eisner}, J., {et~al.} 2018, \apj, 860, 124,
  \dodoi{10.3847/1538-4357/aac6cb}

\bibitem[{{Dullemond} \& {Dominik}(2004)}]{dullemond04a}
{Dullemond}, C.~P., \& {Dominik}, C. 2004, \aap, 417, 159,
  \dodoi{10.1051/0004-6361:20031768}

\bibitem[{{Dullemond} {et~al.}(2012){Dullemond}, {Juhasz}, {Pohl}, {Sereshti},
  {Shetty}, {Peters}, {Commercon}, \& {Flock}}]{dullemond12}
{Dullemond}, C.~P., {Juhasz}, A., {Pohl}, A., {et~al.} 2012, {RADMC-3D: A
  multi-purpose radiative transfer tool}.
\newblock \doeprint{1202.015}

\bibitem[{{Dullemond} \& {Penzlin}(2018)}]{dullemond18a}
{Dullemond}, C.~P., \& {Penzlin}, A.~B.~T. 2018, \aap, 609, A50,
  \dodoi{10.1051/0004-6361/201731878}

\bibitem[{{Dullemond} {et~al.}(2018){Dullemond}, {Birnstiel}, {Huang},
  {Kurtovic}, {Andrews}, {Guzm{\'a}n}, {P{\'e}rez}, {Isella}, {Zhu}, {Benisty},
  {Wilner}, {Bai}, {Carpenter}, {Zhang}, \& {Ricci}}]{dullemond18b}
{Dullemond}, C.~P., {Birnstiel}, T., {Huang}, J., {et~al.} 2018, \apjl, 869,
  L46, \dodoi{10.3847/2041-8213/aaf742}

\bibitem[{{Ercolano} \& {Pascucci}(2017)}]{ercolano17}
{Ercolano}, B., \& {Pascucci}, I. 2017, Royal Society Open Science, 4, 170114,
  \dodoi{10.1098/rsos.170114}

\bibitem[{{Facchini} {et~al.}(2020){Facchini}, {Benisty}, {Bae}, {Loomis},
  {Perez}, {Ansdell}, {Mayama}, {Pinilla}, {Teague}, {Isella}, \&
  {Mann}}]{facchini20}
{Facchini}, S., {Benisty}, M., {Bae}, J., {et~al.} 2020, \aap, 639, A121,
  \dodoi{10.1051/0004-6361/202038027}

\bibitem[{{Flock} {et~al.}(2015){Flock}, {Ruge}, {Dzyurkevich}, {Henning},
  {Klahr}, \& {Wolf}}]{flock15}
{Flock}, M., {Ruge}, J.~P., {Dzyurkevich}, N., {et~al.} 2015, \aap, 574, A68,
  \dodoi{10.1051/0004-6361/201424693}

\bibitem[{{Goldreich} \& {Tremaine}(1980)}]{goldreich80}
{Goldreich}, P., \& {Tremaine}, S. 1980, \apj, 241, 425, \dodoi{10.1086/158356}

\bibitem[{{Guzm{\'a}n} {et~al.}(2018){Guzm{\'a}n}, {Huang}, {Andrews},
  {Isella}, {P{\'e}rez}, {Carpenter}, {Dullemond}, {Ricci}, {Birnstiel},
  {Zhang}, {Zhu}, {Bai}, {Benisty}, {{\"O}berg}, \& {Wilner}}]{guzman18}
{Guzm{\'a}n}, V.~V., {Huang}, J., {Andrews}, S.~M., {et~al.} 2018, \apjl, 869,
  L48, \dodoi{10.3847/2041-8213/aaedae}

\bibitem[{{Hu} {et~al.}(2019){Hu}, {Zhu}, {Okuzumi}, {Bai}, {Wang}, {Tomida},
  \& {Stone}}]{hu2019}
{Hu}, X., {Zhu}, Z., {Okuzumi}, S., {et~al.} 2019, \apj, 885, 36,
  \dodoi{10.3847/1538-4357/ab44cb}

\bibitem[{{Huang} {et~al.}(2018){Huang}, {Andrews}, {P{\'e}rez}, {Zhu},
  {Dullemond}, {Isella}, {Benisty}, {Bai}, {Birnstiel}, {Carpenter},
  {Guzm{\'a}n}, {Hughes}, {{\"O}berg}, {Ricci}, {Wilner}, \&
  {Zhang}}]{huang18c}
{Huang}, J., {Andrews}, S.~M., {P{\'e}rez}, L.~M., {et~al.} 2018, \apjl, 869,
  L43, \dodoi{10.3847/2041-8213/aaf7a0}

\bibitem[{{Huang} {et~al.}(2020){Huang}, {Andrews}, {Dullemond}, {{\"O}berg},
  {Qi}, {Zhu}, {Birnstiel}, {Carpenter}, {Isella}, {Mac{\'\i}as}, {McClure},
  {P{\'e}rez}, {Teague}, {Wilner}, \& {Zhang}}]{huang20}
{Huang}, J., {Andrews}, S.~M., {Dullemond}, C.~P., {et~al.} 2020, \apj, 891,
  48, \dodoi{10.3847/1538-4357/ab711e}

\bibitem[{Hunter(2007)}]{matplotlib}
Hunter, J.~D. 2007, Computing In Science \& Engineering, 9, 90

\bibitem[{{Isella} \& {Turner}(2018)}]{isella2018}
{Isella}, A., \& {Turner}, N.~J. 2018, \apj, 860, 27,
  \dodoi{10.3847/1538-4357/aabb07}

\bibitem[{{Isella} {et~al.}(2018){Isella}, {Huang}, {Andrews}, {Dullemond},
  {Birnstiel}, {Zhang}, {Zhu}, {Guzm{\'a}n}, {P{\'e}rez}, {Bai}, {Benisty},
  {Carpenter}, {Ricci}, \& {Wilner}}]{isella18b}
{Isella}, A., {Huang}, J., {Andrews}, S.~M., {et~al.} 2018, \apjl, 869, L49,
  \dodoi{10.3847/2041-8213/aaf747}

\bibitem[{{Jang-Condell} \& {Turner}(2012)}]{jang-condell12}
{Jang-Condell}, H., \& {Turner}, N.~J. 2012, \apj, 749, 153,
  \dodoi{10.1088/0004-637X/749/2/153}

\bibitem[{{Jang-Condell} \& {Turner}(2013)}]{jang-condell13}
---. 2013, \apj, 772, 34, \dodoi{10.1088/0004-637X/772/1/34}

\bibitem[{{Johansen} {et~al.}(2009){Johansen}, {Youdin}, \&
  {Klahr}}]{johansen09}
{Johansen}, A., {Youdin}, A., \& {Klahr}, H. 2009, \apj, 697, 1269,
  \dodoi{10.1088/0004-637X/697/2/1269}

\bibitem[{Jones {et~al.}(2001--)Jones, Oliphant, Peterson, {et~al.}}]{scipy}
Jones, E., Oliphant, T., Peterson, P., {et~al.} 2001--, {SciPy}: Open source
  scientific tools for {Python}.
\newblock \url{http://www.scipy.org/}

\bibitem[{{Keppler} {et~al.}(2019){Keppler}, {Teague}, {Bae}, {Benisty},
  {Henning}, {van Boekel}, {Chapillon}, {Pinilla}, {Williams}, {Bertrang},
  {Facchini}, {Flock}, {Ginski}, {Juhasz}, {Klahr}, {Liu}, {M{\"u}ller},
  {P{\'e}rez}, {Pohl}, {Rosotti}, {Samland}, \& {Semenov}}]{keppler19}
{Keppler}, M., {Teague}, R., {Bae}, J., {et~al.} 2019, \aap, 625, A118,
  \dodoi{10.1051/0004-6361/201935034}

\bibitem[{{Klahr} \& {Bodenheimer}(2003)}]{klahr03}
{Klahr}, H.~H., \& {Bodenheimer}, P. 2003, \apj, 582, 869,
  \dodoi{10.1086/344743}

\bibitem[{{Lin} \& {Papaloizou}(1979)}]{lin79}
{Lin}, D.~N.~C., \& {Papaloizou}, J. 1979, \mnras, 186, 799

\bibitem[{{Liu}(2019)}]{Liu2019}
{Liu}, H.~B. 2019, \apjl, 877, L22, \dodoi{10.3847/2041-8213/ab1f8e}

\bibitem[{{Long} {et~al.}(2020){Long}, {Pinilla}, {Herczeg}, {Andrews},
  {Harsono}, {Johnstone}, {Ragusa}, {Pascucci}, {Wilner}, {Hendler},
  {Jennings}, {Liu}, {Lodato}, {Menard}, {van de Plas}, \& {Dipierro}}]{long20}
{Long}, F., {Pinilla}, P., {Herczeg}, G.~J., {et~al.} 2020, \apj, 898, 36,
  \dodoi{10.3847/1538-4357/ab9a54}

\bibitem[{{Mac{\'\i}as} {et~al.}(2019){Mac{\'\i}as}, {Espaillat}, {Osorio},
  {Anglada}, {Torrelles}, {Carrasco-Gonz{\'a}lez}, {Flock}, {Linz}, {Bertrang},
  {Henning}, {G{\'o}mez}, {Calvet}, \& {Dent}}]{macias19}
{Mac{\'\i}as}, E., {Espaillat}, C.~C., {Osorio}, M., {et~al.} 2019, \apj, 881,
  159, \dodoi{10.3847/1538-4357/ab31a2}

\bibitem[{{Morbidelli}(2020)}]{Morbidelli2020}
{Morbidelli}, A. 2020, \aap, 638, A1, \dodoi{10.1051/0004-6361/202037983}

\bibitem[{{Nelson} {et~al.}(2013){Nelson}, {Gressel}, \&
  {Umurhan}}]{Nelson2013}
{Nelson}, R.~P., {Gressel}, O., \& {Umurhan}, O.~M. 2013, \mnras, 435, 2610,
  \dodoi{10.1093/mnras/stt1475}

\bibitem[{{Pinte} {et~al.}(2018){Pinte}, {M{\'e}nard}, {Duch{\^e}ne}, {Hill},
  {Dent}, {Woitke}, {Maret}, {van der Plas}, {Hales}, {Kamp}, {Thi}, {de
  Gregorio-Monsalvo}, {Rab}, {Quanz}, {Avenhaus}, {Carmona}, \&
  {Casassus}}]{pinte18}
{Pinte}, C., {M{\'e}nard}, F., {Duch{\^e}ne}, G., {et~al.} 2018, \aap, 609,
  A47, \dodoi{10.1051/0004-6361/201731377}

\bibitem[{{Rosotti} {et~al.}(2021){Rosotti}, {Ilee}, {Facchini}, {Tazzari},
  {Booth}, {Clarke}, \& {Kama}}]{rosotti21}
{Rosotti}, G.~P., {Ilee}, J.~D., {Facchini}, S., {et~al.} 2021, \mnras, 501,
  3427, \dodoi{10.1093/mnras/staa3869}

\bibitem[{{Sato} {et~al.}(2016){Sato}, {Okuzumi}, \&
  {Ida}}]{2016A&A...589A..15S}
{Sato}, T., {Okuzumi}, S., \& {Ida}, S. 2016, \aap, 589, A15,
  \dodoi{10.1051/0004-6361/201527069}

\bibitem[{{Takahashi} \& {Muto}(2018)}]{takahashi18}
{Takahashi}, S.~Z., \& {Muto}, T. 2018, \apj, 865, 102,
  \dodoi{10.3847/1538-4357/aadda0}

\bibitem[{{Van Der Walt} {et~al.}(2011){Van Der Walt}, {Colbert}, \&
  {Varoquaux}}]{numpy}
{Van Der Walt}, S., {Colbert}, S.~C., \& {Varoquaux}, G. 2011, ArXiv e-prints.
\newblock \doarXiv{1102.1523}

\bibitem[{{Weidenschilling}(1977)}]{1977MNRAS.180...57W}
{Weidenschilling}, S.~J. 1977, \mnras, 180, 57, \dodoi{10.1093/mnras/180.2.57}

\bibitem[{{Youdin} \& {Lithwick}(2007)}]{youdin2007}
{Youdin}, A.~N., \& {Lithwick}, Y. 2007, \icarus, 192, 588,
  \dodoi{10.1016/j.icarus.2007.07.012}

\bibitem[{{Zhang} {et~al.}(2015){Zhang}, {Blake}, \& {Bergin}}]{zhang2015}
{Zhang}, K., {Blake}, G.~A., \& {Bergin}, E.~A. 2015, \apjl, 806, L7,
  \dodoi{10.1088/2041-8205/806/1/L7}

\bibitem[{{Zhu} {et~al.}(2012){Zhu}, {Nelson}, {Dong}, {Espaillat}, \&
  {Hartmann}}]{zhu12}
{Zhu}, Z., {Nelson}, R.~P., {Dong}, R., {Espaillat}, C., \& {Hartmann}, L.
  2012, \apj, 755, 6, \dodoi{10.1088/0004-637X/755/1/6}

\bibitem[{{Zhu} {et~al.}(2019){Zhu}, {Zhang}, {Jiang}, {Kataoka}, {Birnstiel},
  {Dullemond}, {Andrews}, {Huang}, {P{\'e}rez}, {Carpenter}, {Bai}, {Wilner},
  \& {Ricci}}]{zhu19b}
{Zhu}, Z., {Zhang}, S., {Jiang}, Y.-F., {et~al.} 2019, \apjl, 877, L18,
  \dodoi{10.3847/2041-8213/ab1f8c}

\end{thebibliography}
\end{document}